\documentclass[prl,amsmath,amssymb,reprint]{revtex4-2}
\usepackage{graphicx}
\usepackage{mathtools}
\usepackage{hyperref}
\usepackage{mhchem} 
\usepackage{natbib}

\begin{document}

\title{\textbf{Critical quantum metrology beyond adiabaticity in collectively pumped superradiance}}
\author{Yoav Shimshi}
\affiliation{Department of Chemical \& Biological Physics, Weizmann Institute of Science, Rehovot 7610001, Israel}
\author{Ephraim Shahmoon}
\affiliation{Department of Chemical \& Biological Physics, Weizmann Institute of Science, Rehovot 7610001, Israel}
\date{\today}
\begin{abstract}
Critical metrology relies on the high sensitivity of systems to parameter changes near a phase transition to extract information with high precision. Such methods usually require working under adiabatic conditions, which implies long preparation times due to critical slowing down. Here instead we demonstrate how favorable sensitivity can be maintained for preparation rates exceeding the adiabatic limit by exploiting knowledge of the non-equilibrium dynamics of the system. Our work focuses on a model of collective dissipation (superradiance) balanced by collective pumping, analyzing the sensitivity around a first order dissipative phase transition. We devise a sensing protocol based on characterizing the hysteresis formed beyond adiabatic conditions, finding the sensing uncertainty $r^{\alpha}/N$, exhibiting a favorable Heisenberg-like scaling with the particle number $N$ and only a weak power-law dependence $\alpha\approx0.17$ on the preparation rate $r$. We discuss the implementation of the model and potential applications to metrology within cavity QED setups.
\end{abstract}


\maketitle


Critical metrology exploits the great sensitivity of a stable state near criticality to changes in external parameters for precision measurements \cite{Montenegro2025}. In quantum systems, the stable state can be either the ground state of a many-body Hamiltonian near a quantum phase transition \cite{Frerot2018,Salvia2023,Garbe2020}, or the steady state of a driven quantum open system near a dissipative phase transition \cite{FernandezLorenzo2017,Marzolino2014,Pavlov2023,Heugel2019,Macieszczak2016,Ilias2022,Ding2022,Montenegro2023}. Taking both preparation time and system size as resources, it was theoretically shown in both cases that the Heisenberg limit of sensitivity can be reached \cite{Invernizzi2008,Rams2018,Ivanov2013,Lu2024}.

Critical protocols typically require long times to adiabatically bring the system near criticality \cite{Rams2018}. In quantum phase transitions the adiabatic conditions are dictated by an energy gap, while in dissipative transitions the Liouvillian gap gives the relevant timescale. In both cases the gap typically closes at the transition, leading to critical slowing down of the dynamics \cite{Garbe2020,Lue2024}. Thus, metrology at the stable state requires either prolonged preparation times or the application of non-trivial driving to combat non-adiabatic effects \cite{Gietka2021,Gietka2022,Roland2002,DelCampo2019}.

While much work on analyzing the sensitivity of  critical protocols under adiabatic conditions is being done, less attention is given to the possibility of metrological benefit beyond it \cite{Montenegro2025}. It is thus interesting to ask whether Heisenberg-like scaling can be preserved without using the stable adiabatic state, instead extracting information directly from the non-adiabatic dynamics of the system. To address this question we consider a model where spins (``atoms") undergo a dissipative phase transition due to a competition between collective dissipation (superradiance) and collective pumping. At steady state, it displays a first order transition at a critical pumping rate, with a susceptibility scaling linearly with the atom number $N$, implying the Heisenberg-like scaling $1/N$ for metrology under adiabatic, steady-state conditions. Going well beyond adiabatic conditions, we then study the full dynamics and observe the response to a changing pumping rate around the critical point.

Analyzing the system's hysteresis behavior away from adiabaticity, we are able to construct a protocol to find its critical point, accounting for finite resources such as the atom number $N$ and preparation rate $r$. Our key finding is that the protocol's error in estimating the critical point scales as $r^{\alpha}/N$, with $\alpha\approx0.38\text{ or }0.17$, depending on the protocol. This means that the estimation of the critical point, which
translates to metrology of the pump rate, retains the Heisenberg-like scaling
$1/N$, with performance only weakly affected by the non-adiabatic
dynamics induced by the preparation rate. Importantly, this shows that knowledge of a system's hysteresis behavior can be used to extend the benefits of critical metrology much beyond the adiabatic limit. We relate our results to realistic schemes of superradiance in current cavity QED setups, discussing possible applications in metrology of various quantities of interest.

\begin{figure}[t!]
  \centering
  \includegraphics[width=\columnwidth]{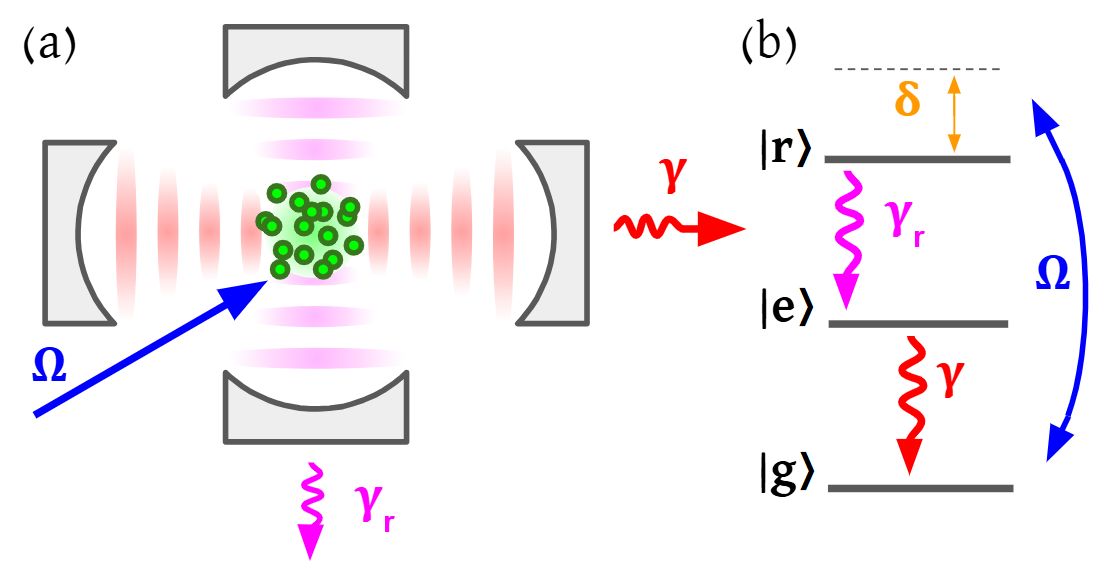}
  \caption{\small{\label{fig (a):schematics} Proposed realization of collectively pumped superradiance, Eq. (\ref{eq:effective master equation}). (a) An atomic ensemble coupled to two independent cavity modes (either of two distinct cavities or of the same cavity \cite{supplement_arxiv}). One mode mediates collective
  decay at rate $\gamma$, while the other mode, together with external laser light (Rabi frequency $\Omega$ and detuning $\delta$), generates collective pumping at a rate $w=\gamma_r(\Omega/\delta)^2$. (b) Level structure of each atom.}}\label{fig (b): System setup}
\end{figure}
\nocite{Reiter2012,Bohnet2012,Thompson1998,Tan2019,Keesling2019}

\emph{Model and realization}.---We consider $N$ two-level atoms undergoing collective dissipation and pumping, as described by the following master equation for the many-atom density matrix $\hat{\rho}$,
\begin{equation}
\dot{\hat{\rho}}=\gamma\mathcal{D}_{\hat{J}_{-}}[\hat{\rho}]+w\mathcal{D}_{\hat{J}_{+}}[\hat{\rho}]\label{eq:effective master equation},
\end{equation}
with $\mathcal{D}_{\hat{A}}[\hat{\rho}]\equiv\hat{A}\hat{\rho}\hat{A}^{\dagger}-\frac{1}{2}\{ \hat{A}^{\dagger}\hat{A},\hat{\rho}\}$ denoting the Lindblad-form irreversible process governed by a jump operator $\hat{A}$. The first term, with jump operator $\hat{J}_{-}=\sum_{n=1}^{N}|g_{n}\rangle\langle e_{n}|$ ($|g_{n}\rangle$ and $|e_{n}\rangle$ denoting the ground and excited levels of atom $n$), accounts for collective radiation at rate $\gamma$, while the second term, with jump operator $\hat{J}_{+}=\hat{J}_{-}^{\dagger}$, describes incoherent, collective pumping of the atoms at rate $w$.
Notably, the dynamics of Eq. (\ref{eq:effective master equation}) conserve the total spin number $j$ of the atomic pseudo-spins. Assuming an initial fully unpopulated state, we have $j=N/2$ and the dynamics are confined to the Hilbert space of $N+1$ Dicke states $|m\rangle$, with $m=-N/2,...,N/2$ \cite{Dicke1954,Gross1982,Kirton2019}.

The above master equation can be realized in cavity QED experiments by simultaneously inducing superradiant emission in two transitions of a three-level pumping scheme, a process which we term ``collective pumping" (Fig. \ref{fig (a):schematics}). The derivation of Eq. (\ref{eq:effective master equation}), based on the adiabatic eliminations of the cavity modes and higher atomic levels, is outlined in the End Matter section and in \cite{supplement_arxiv}. We find an effective pumping rate $w=\gamma_{r}(\Omega/\delta)^{2}$, with $\Omega$ and $\delta$ being, respectively, the Rabi frequency and detuning of the pump laser to the third level. Here $\gamma\equiv C\gamma_{eg}$ and $\gamma_{r}\equiv C_{r}\gamma_{re}$ are the Purcell-enhanced single-atom decays, assuming large enough cooperativities $C,C_{r}\gg1$ for which the individual off-axis decays $\gamma_{eg},\gamma_{er}$
become negligible at experimental timescales. The particular two-level description of Eq.(\ref{eq:effective master equation}) is obtained at off resonance,
$\delta\gg\sqrt{N}\Omega$, wherein all collective states involving
the third level $|r\rangle$ are adiabatically eliminated.
The validity conditions of Eq. (\ref{eq:effective master equation}) and their realization in actual systems are
summarized and discussed further below.

\emph{Steady state solution}.---
Looking at Eq. (\ref{eq:effective master equation}), we note that the pumping term is mathematically analogous to the effect of heat flowing into the system from a finite temperature reservoir. In particular, we identify an effective ``inverse temperature" $\beta\equiv-\ln\left(w/\gamma\right)$, thus anticipating that the steady state would be equivalent to thermal Boltzmann statistics. Indeed, we find
that Eq. (\ref{eq:effective master equation}) has the unique steady state \cite{supplement_arxiv}
\begin{align}
\hat{\rho}_{\text{s}} & =\tfrac{1}{Z\left(\beta\right)}e^{-\beta\hat{J}_{z}}\ ,\ Z\left(\beta\right)=\frac{\sinh\left[\left(N+1\right)\beta/2\right]}{\sinh\left(\beta/2\right)},\label{eq:Steady state density matrix}
\end{align}
with the population inversion operator $\hat{J}_{z}=\sum_{n=1}^{N}(|e_{n}\rangle\langle e_{n}|-|g_{n}\rangle\langle g_{n}|)/2$ taking the role of ``energy". We can derive exact
expressions for moments of $\hat{J}_{z}$ using
the partition function, such as $\langle\hat{J}_{z}\rangle=-\partial_{\beta}\ln Z\left(\beta\right)$. For $N\gg1$ and $\left|w/\gamma-1\right|\ll1$ we obtain
\begin{align}
\langle\hat{J}_{z}\rangle & =\frac{N/2}{\tanh\left(N\left(w/\gamma-1\right)/2\right)}-\frac{1}{w/\gamma-1}\label{eq:mean population}\\
 & \overset{N\rightarrow\infty}{\rightarrow}\begin{cases}
\frac{N}{2}\textnormal{sign}\left(w/\gamma-1\right) & \left|w/\gamma-1\right|\gg\frac{1}{N}\\
\frac{N^{2}}{12}\left(w/\gamma-1\right) & \left|w/\gamma-1\right|\ll\frac{1}{N}
\end{cases}.\nonumber
\end{align}
\begin{figure}[t!]
  \centering
  \includegraphics[width=\columnwidth]{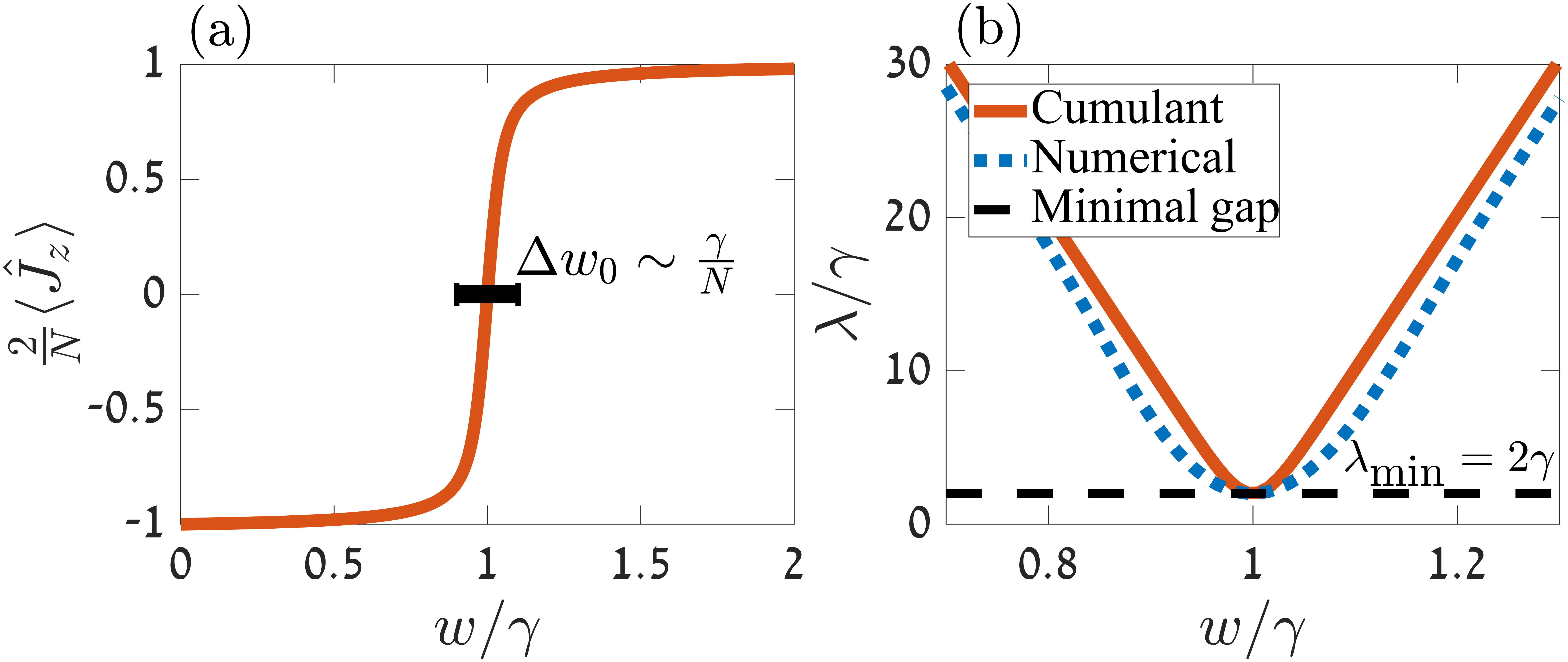}
  \caption{\label{fig (a): mean inversion}(a) The mean population inversion at steady state, $\langle\hat{J}_z\rangle$ from Eq. (\ref{eq:mean population}), as a function of pumping rate $w$ exhibits a sharp transition from depopulated to populated
  state at the critical value $w=\gamma$. (b) Relaxation rate $\lambda$ as a function of pumping
  $w$. Good agreement between analytical cumulant-expansion results of Eq. (\ref{eq:correlation time}) (red) and the numerical calculation of Liouville gap
  of Eq. (\ref{eq:effective master equation}) (dotted blue) is observed, yielding a finite minimal gap
  $\lambda_{\textnormal{min}}=2\gamma$ at the critical point $w=\gamma$
  (dashed black). While the plots shown here are for $N=100$, we verify that the gap $\lambda_{\textnormal{min}}=2\gamma$ is indeed independent of system size $N$, allowing to reach the steady state at a finite time $\gtrsim \gamma^{-1}$ even at the thermodynamic limit.}\label{fig (b): relaxation rate}
\end{figure}
In the thermodynamic limit, $N\rightarrow\infty$, the system exhibits
a sharp jump at a critical pumping rate $w=\gamma$ from a fully depleted
($w<\gamma$) to a fully populated state ($w>\gamma$), as clearly
seen in the exact solution plotted in Fig. \ref{fig (a): mean inversion}a. This abrupt switching between the two states originates in the nonlinear pull from the competing
dissipation and pumping processes in Eq. (\ref{eq:effective master equation}), which are exactly balanced at the half population crossing point $w=\gamma$. The width of the crossing region is of order $\gamma/N$, shrinking to zero
as $N\rightarrow\infty$.

\emph{Metrology with the stable state}.---Evidently, near the transition measurements of the mean population
are highly sensitive to changes in the pumping rate $w$. One might
then imagine a procedure where we scan $w$ adiabatically and locate the critical rate $w=\gamma$, at which the population
jumps, to an accuracy of $\Delta w_0\sim\gamma/N$. This can be achieved by measuring $\hat{J}_z$ either by continuous monitoring of off-cavity-axis fluorescence or by using shelving techniques \cite{Song2025}. We discuss the viability of these measurement methods, along with a method based on detection of on-cavity axis emission, in the End Matter.

In either case, the resulting small uncertainty in measuring $w$ can then be exploited for precise estimation of intrinsic atomic parameters or applied fields that determine $w$ (End Matter). Extracted from population measurements around
the crossing point $w=\gamma$ (where $\beta\approx1-w/\gamma$), this uncertainty can be quantified by the slope of $\langle\hat{J}_{z}\rangle$ and the fluctuations $\text{Var}[\hat{J}_{z}]=\partial_{\beta}^{2}\ln Z\left(\beta\right)=-\partial_{\beta}\langle\hat{J}_{z}\rangle$, as per the error propagation formula, obtaining
\begin{equation}
\Delta w_0\approx\frac{\sqrt{\text{Var}[\hat{J}_{z}]}}{|\partial_{w}\langle\hat{J}_{z}\rangle|}=\frac{\sqrt{12}\gamma}{N}.\label{eq: Steady state sensitivity}
\end{equation}
The above result is valid when the system has reached steady state. To estimate the time needed to approach stability we must calculate the system relaxation rate $\lambda$ as a function of pumping $w$. This can be done using the two-time
correlation function $C_{zz}(\tau)=\langle\hat{J}_z(t+\tau)\hat{J}_z(t)\rangle-\langle\hat{J}_z(t+\tau)\rangle\langle\hat{J}_z(t)\rangle$,
which exhibits exponential scaling $e^{-\lambda\tau}$
at long times. Using the quantum regression theorem \cite{Swain1981} and neglecting third order cumulants \cite{Kubo1962} we derive \cite{supplement_arxiv}
\begin{equation}
\partial_{\tau}C_{zz}\left(\tau\right)\approx-\lambda\left(w\right)C_{zz}\left(\tau\right),\label{eq:population-population correlations}
\end{equation}
with the relaxation rate given by
\begin{align}
\lambda\left(w\right) & =\frac{N(w-\gamma)}{\tanh(N/2(w/\gamma-1))}\label{eq:correlation time}\\
 & \approx\begin{cases}
N\left|w-\gamma\right| & \left|w/\gamma-1\right|\gg\frac{1}{N}\\
2\gamma & \left|w/\gamma-1\right|\ll\frac{1}{N}
\end{cases}.\nonumber
\end{align}
In Fig. \ref{fig (b): relaxation rate}b we plot the rate from Eq. (\ref{eq:correlation time}), compared to that obtained by a numerical calculation
of the lowest non-vanishing eigenvalue (Liouville gap) of the rate equation for populations $\left\langle m\right|\hat{\rho}\left|m\right\rangle $
obtained from Eq. (\ref{eq:effective master equation}) \cite{supplement_arxiv}. Very good agreement is observed for relevant regions, indicating a decrease of the relaxation rate as one approaches the transition --- this is the phenomenon of critical slowing down \cite{Binder1984,Berglund1998}.
Interestingly, we obtain that the minimal rate reached at
the critical point $w=\gamma$ has a finite value, $\lambda_{\textnormal{min}}=2\gamma$,
even at the limit $N\rightarrow\infty$. This means that the system always reaches steady state at a finite time, ensuring the possibility for adiabatic conditions and hence the favorable $1/N$ scaling from (\ref{eq: Steady state sensitivity}) regardless of system size.

\begin{figure}[t!]
  \centering
\includegraphics[width=\columnwidth]{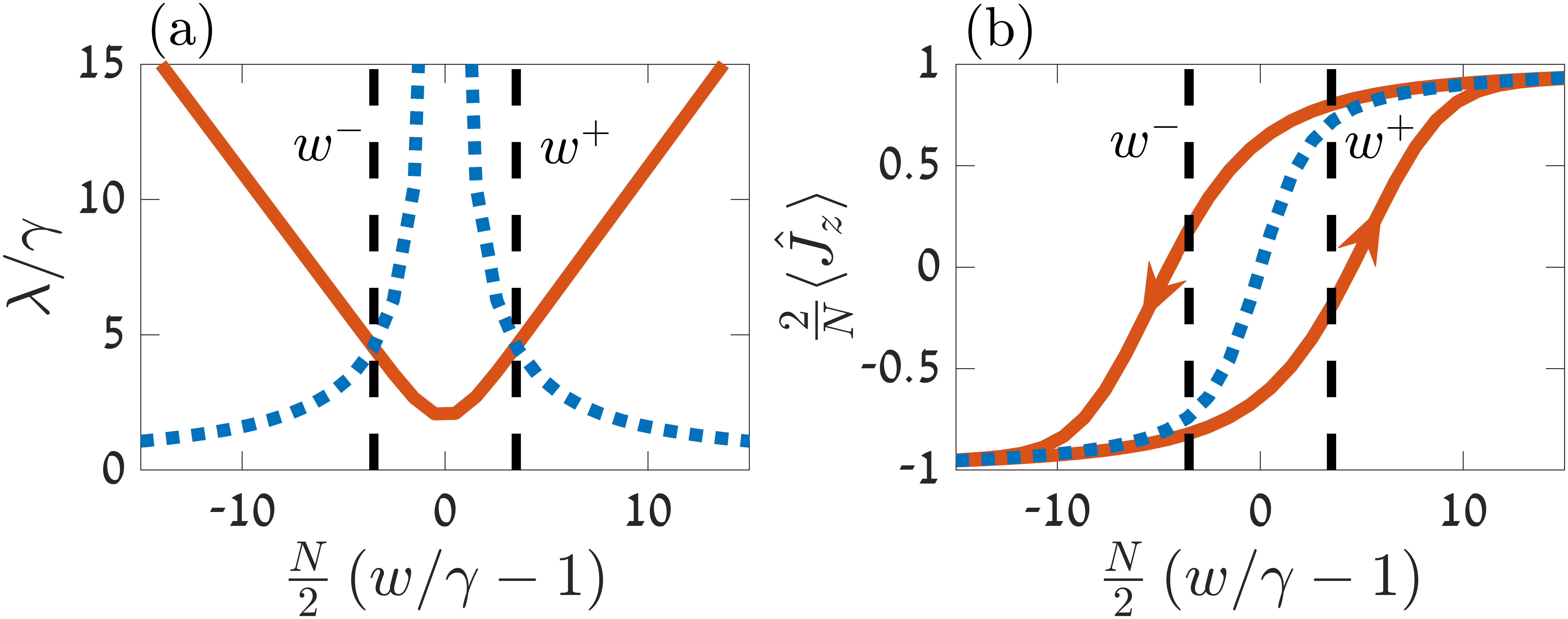}
  \caption{\label{fig (a):relaxation vs scanning}(a) Relaxation rate $\lambda$ (red) and
normalized scanning rate $\dot{w}/\left|w-\gamma\right|\propto 1/|w-\gamma|$ of the pumping strength $w$ (dashed blue) as a function of $w$, for constant scanning rate  $r\equiv \dot{w}/\Delta w_0=16\gamma$ and $N=100$. Dashed horizontal lines mark the boundaries $w^{\pm}$ within which the scanning rate is too fast for the system to adibatically follow the steady state solution.
 \label{fig (b):hysteresis} (b) A hysteresis loop of the mean inversion $\langle\hat{J}_z\rangle$ (red) is formed around the steady state solution (\ref{eq:mean population}) (dashed blue) due to the dynamics induced by the scanning rate. The width of the hysteresis grows proportionally to the region of nonadiabaticity deduced from (a).}
\end{figure}

\emph{Non-adiabatic dynamics and hysteresis}.---
While the gap is always finite, it might still be very small in some particular realizations, motivating us to study the behavior of the system outside of adiabatic conditions. Our key result here is that even beyond adiabaticity the favorable $1/N$ scaling may be maintained by properly accounting for the system's non-equilibrium dynamics. For concreteness, suppose we attempt to prepare the system at a state corresponding to a specific pumping by scanning $w$ from $0$ to the target value at a rate $\dot{w}=\Delta w_{0}\cdot r$, with $1/r$ being the typical time to pass the transition point. If the scanning is done faster than the rate at which the system can respond, we expect the measured $\langle\hat{J}_{z}\rangle$ to lag behind the true steady state, leading to a value lower than expected. By reversing the direction of scanning we accordingly get a value higher then expected, leading to a hysteresis-like curve when plotting the results against pumping (Fig. \ref{fig (b):hysteresis}b).

One way to characterize the hysteresis shape is to define the mean $w^{\pm}$ and uncertainty $\Delta w^{\pm}$ of the points at which the hysteresis shape crosses zero for each direction of scanning: The difference of these crossing points quantifies the hysteresis width, while $\Delta w^{\pm}$, defined in analogy to Eq. (\ref{eq: Steady state sensitivity}), are the sensitivities of estimating the crossing points from measurements of $\hat{J}_{z}$. We assume power law scaling for both these quantities, and write:
\begin{align}
w^{+}-w^{-} & \sim\Delta w_0\left(r/\gamma\right)^{\eta},\label{eq: 8.hysteresis exponents}\\
\Delta w^{\pm} & \sim\Delta w_0\left(r/\gamma\right)^{\varepsilon}.\nonumber
\end{align}
Ultimately, the effectiveness of estimating the critical point $w=\gamma$
must depend on the $r$ scaling of the above quantities, and thus
we are interested in predicting the exponents appearing in (\ref{eq:
8.hysteresis exponents}). Beginning with the exponent $\eta$, we note that the half population points
$w^{\pm}$ roughly correspond to the
boundaries of the nonadiabatic region in Fig. \ref{fig
(b):hysteresis}a. In this region the instantaneous scanning rate
$\sim\frac{\dot{w}}{|w-\gamma|}$ should be higher than the natural
relaxation rate $\lambda$, given by (\ref{eq:correlation time}). Equating
these two rates should then give the boundary of the nonadiabatic
region and hence the hysteresis width, finding
\begin{equation}
w^{+}-w^{-}\sim\Delta w_0\times\begin{cases}
r/\gamma & r\ll\gamma\\
\sqrt{r/\gamma} & r\gg\gamma.
\end{cases}\label{eq: 9.hysteresis width}
\end{equation}
In the nonadiabatic regime $r\gg\gamma$, this yields the exponent $\eta\approx 1/2$, originated in the linear scaling $\lambda\propto |w-\gamma|$ of the relaxation rate with pumping away from criticality, at $|w-\gamma|\gg \Delta w_0$ (see also \cite{supplement_arxiv}).
The scaling exponent $\varepsilon$ of crossing point uncertainties is harder
to predict analytically, and we resort to a numerical simulation of the preparation process.

To this end, we perform a direct numerical integration of (\ref{eq:effective master equation}) with a time dependent pumping $w(t)=\Delta w_0 r t$, obtaining the density matrix as a function of time. Plotting $\langle\hat{J}_{z}(t)\rangle$ as a function of $w\left(t\right)$
for positive and negative $r$ then gives us a hysteresis loop, as shown
in Fig. \ref{fig (a):relaxation vs scanning}b for $r=16\gamma$ and $N=100$. From this we extract
the crossing points and their respective uncertainties defined in analogy to Eq. (\ref{eq:
Steady state sensitivity}), and by repeating the procedure for different
$r$ we obtain a plot of $w^{+}-w^{-}$ and $\Delta w^{\pm}$ as a
function of $r$, as shown in Fig. \ref{fig (a):numerical
simulation log fit}.

For $r\ll\gamma$ the width of the hysteresis scales linearly with
$r$ in accordance with (\ref{eq: 9.hysteresis width}),
while the uncertainties around each crossing point approach the steady-state
uncertainty found in (\ref{eq: Steady state sensitivity}). This recovers adiabatic behavior expected at sufficiently slow scanning rates, noting that the hysteresis width becomes smaller than the uncertainty and is hence effectively washed out by the noise.
For $r\gg\gamma$ we see the expected power law scaling in both quantities,
with $\eta\approx0.58$ and $\varepsilon\approx0.38$. The slight deviation in $\eta$, compared to our analytical prediction $\eta\approx1/2$ from (\ref{eq: 9.hysteresis width}), is attributed
to the existence of additional timescales in the long-time behavior
of Eq. (\ref{eq:effective master equation}), which perturb the relaxation dynamics from the assumed
purely exponential decay at a single rate $\lambda$ \cite{supplement_arxiv}.
We have also compared the curves achieved
for different atom numbers in the range $N\in\{30,100\}$, and found that upon rescaling
the widths to $\Delta w_0$ they all collapse on each other,
indicating that the obtained exponents are independent
of atom number \cite{supplement_arxiv}.

\emph{Metrology beyond adiabaticity}.---We now show how to exploit our knowledge of the hysteresis shape
to provide an estimate for the critical point $w=\gamma$ in nonadiabatic
conditions, $r\gg\gamma$. To do this, we define
the new estimate to be the average $(w^{+}+w^{-})/2$ of the two crossing
points. This presents an unbiased estimator thanks to a symmetry of our system to a simultaneous inversion
of population and the scanning rate \cite{supplement_arxiv}, which guarantees that
the two crossing points evolve symmetrically around the original transition. The resulting single-shot error for
estimating the transition is then
\begin{equation}
\Delta w(N,r)\approx\Delta w^{\pm}/\sqrt{2}\sim\Delta w_0\times\begin{cases}
1 & r\ll\gamma\\
(r/\gamma)^{\varepsilon} & r\gg\gamma
\end{cases}.\label{eq:Single shot sensitivity}
\end{equation}
Remarkably, this shows that even above the optimal adiabatic rate $r\sim\gamma$ the sensitivity retains its $\Delta w_0\propto 1/N$ scaling, with a moderate widening factor $\propto(r/\gamma)^{0.38}$.

\begin{figure}[t!]
  \centering
\includegraphics[width=\columnwidth]{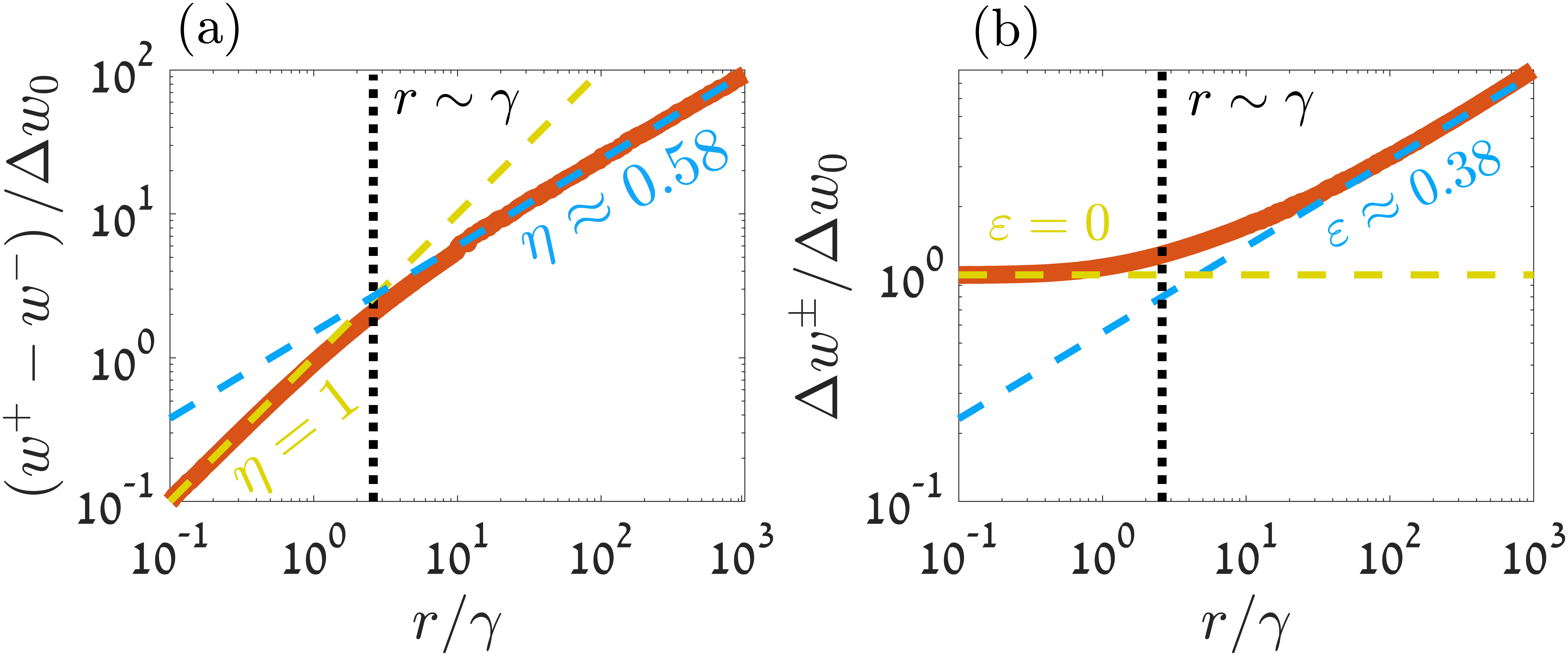}
  \caption{\label{fig (a):numerical simulation log fit}
  Scaled hysteresis width $(w^{+}-w^{-})/\Delta w_{0}$ and crossing
point errors $\Delta w^{\pm}/\Delta w_{0}$ vs. scanning rate $r$,
obtained numerically for $N=500$.
(a) Scaled hysteresis width. Dashed asymptotes indicate power-law
behavior, switching from a power $\eta=1$ to a power $\eta\approx0.58$
around a critical scanning rate $r\sim\gamma$, similar to the prediction
of Eq. (\ref{eq: 9.hysteresis width}). Deviation
from the $\eta=0.5$ prediction for $r>\gamma$ originates in the
existence of multiple timescales \cite{supplement_arxiv}. 
Notably, these results are independent of $N$, as verified by the collapse of curves for different $N$ on each other \cite{supplement_arxiv}.
(b) Scaled crossing errors. The plots for $\Delta w^{+}$ and $\Delta w^{-}$
are identical, and so the red curve represents the results for both
of them. The asymptotes show a transition from the equilibrium behavior
$\Delta w^{\pm}=\Delta w_{0}$ below $r\sim\gamma$ to a power law
scaling with exponent $\varepsilon\approx0.38$. \label{fig (b):numerical simulation different N}}\label{Hysteresis scaling law}
\end{figure}

The sensitivity can be improved even further by considering repeated measurement shots within a total interrogation time $T$. This time is set by any competing decay process, after which the system has to be re-initialized: in the cavity QED realization, this may correspond to off-cavity-axis decay $\gamma_{eg}$ \cite{supplement_arxiv}. Optimally, we should continuously gather information from the system and incorporate it into our estimation of the transition - this is possible by collecting the light coming out of the cavity either in the on-axis or off-axis direction \cite{supplement_arxiv}. We can then periodically scan the pumping around our working point at a given scanning rate $r$, and construct from our measurement an ensemble of hysteresis curves of $\langle\hat{J}_{z}\rangle$ versus $w$ (whose average yields the hysteresis in Fig. \ref{fig (b):hysteresis}b).

Since we want to capture the full width of the transition region, given by $\Delta w_{0}$ or the hysteresis width in the adiabatic or nonadiabatic regimes, respectively, we should set the scan range to be $\Delta w_{\text{scan}}\sim A\cdot\max\{\Delta w_{0},w^{+}-w^{-}\}$. Here, the dimensionless amplitude $A$ is chosen to be of order $1$, which can be achieved by iteratively optimizing the scanning window. At the edge of the sweep range the system’s relaxation is very fast (Eq. (\ref{eq:correlation time})), meaning that each sweep is essentially statistically independent. The total error is then given by the single shot sensitivity from Eq. (\ref{eq:Single shot sensitivity}) divided by the square root of the number of periods $N_{\text{scan}}\sim\dot{w}T/\Delta w_{\text{scan}}$, yielding
\begin{align}
\Delta w(N,r,A,T)\sim\Delta w_0\sqrt{\frac{A}{\gamma T}}\times\begin{cases}
(r/\gamma)^{-1/2} & r\ll\gamma\\
(r/\gamma)^{\varepsilon-\frac{1-\eta}{2}} & r\gg\gamma
\end{cases}.\label{eq: 10.total uncertainty}
\end{align}
Again we see that the $1/N$ scaling is preserved above adiabatic conditions, with a significantly improved scaling with the scanning rate $r$ compared to the single-scan result: we obtain a scaling exponent of $\varepsilon-\frac{1-\eta}{2}\approx0.17$, much weaker than the $\varepsilon\approx0.38$ scaling law from Eq. (\ref{eq:Single shot sensitivity}). The additional pre-factor $\sqrt{A/(\gamma T)}$ depends on the interrogation time $T$: in our case, where $T\sim \gamma_{eg}^{-1}=C\gamma^{-1}$, this only gives a moderate contribution $\sqrt{A/C}$ decreasing with the cooperativity $C$.

\emph{Realization and applications}.---Let us return to the question of realization. An ensemble of three-level atoms such as in Fig. \ref{fig (b): System setup}a can be made to interact with two independent cavity modes either by trapping it between two different cavities \cite{Leroux2010,SchleierSmith2010,Yan2023,Hayn2011,Hayn2012,Haake1996}, or by placing the atoms in particular locations along a single cavity where two modes spatially overlap \cite{supplement_arxiv}. Assuming the cavity's leakage rates to be much faster than the atomic ensemble timescale, we obtain a master equation for the atoms, with cavity-enhanced collective decay rates $\gamma=C\gamma_{eg}$ and $\gamma_{r}=C_{r}\gamma_{re}$ ($C,C_{r}\gg1$), and neglecting the slower off-cavity-axis individual decays during the experiment time $T<\gamma_{eg}^{-1}$. To achieve collective pumping we further adiabatically eliminate the third levels $|r\rangle$ by assuming their negligible population, $\sim N\Omega^{2}/\delta^{2}\ll 1$ (End Matter). Around the critical point $w\equiv\gamma_{r}\Omega^{2}/\delta^{2}=\gamma$ this condition implies $N\gamma\ll\gamma_{r}$, meaning that we need to choose particular transitions with vastly different decay timescales to make our model feasible. This is possible e.g. for the hyperfine transition of \ce{^{87}Sr}, for which we obtain $\gamma_{r}/\gamma\sim10^{8}$ with the relevant levels $|g\rangle\equiv(5s^{2})1S0$, $|e\rangle\equiv(5s5p)3P0$
and $|r\rangle\equiv(5s7s)3S1$ \cite{supplement_arxiv}, and where the $|g\rangle$-$|r\rangle$ coupling could be achieved by a two-photon transition through the level $(5s5p)3P1$ \cite{Lu2024,Sansonetti2010}. While the condition implies an ultimate bound on the number of atoms $N\gg\gamma_{r}/\gamma\sim10^{8}$, this number is in practice well beyond typical atom numbers used in cavity QED experiments with strontium \cite{Norcia2016,Norcia2016a}.

So far, we showed how the critical transition point $w=\gamma_{r}(\Omega/\delta)^{2}=\gamma$ can be measured to a precision of $O(1/N)$. By error propagation this favorable $1/N$ sensitivity can be exploited to measure any quantity that enters into the critical pumping value. One example is the estimation of small changes in the cavity length (e.g. in relation to measurement of external forces), which affect the cavity-enhanced decay $\gamma$ and hence shifts the critical point (End Matter). Another interesting application is the determination of the atom's transition lifetimes, where ideally the error in the fast transition $\gamma_{r}$ is improved to the level of the slow one $\gamma$ through determination of $(\Omega/\delta)^{2}=\gamma/\gamma_{r}$. Current precision standards for lifetime measurements are relatively modest \cite{Tanner1994,Traebert2024,Monier2010} (e.g. $\gtrsim0.1\%$ for strontium \cite{Sansonetti2010}), so this method might be useful for applications that require knowledge of atomic lifetimes, such as fundamental physics tests \cite{Nemouchi2009} or astrophysical spectra \cite{Pickering2011}.

\emph{Discussion}.---In summary, we show that by studying non-equilibrium dynamics it may be possible to devise methods, e.g. based on hysteresis, to enjoy critical metrology beyond the adiabatic limit. The presented protocol is robust to changes in preparation rate, and is relatively simple to implement, compared to e.g. shortcut to adiabaticity methods whose complexity increases with system size \cite{DelCampo2019}. Our analysis, while applicable in cavity QED settings, is not restricted to the present realization, and can be generalized to any situation requiring scanning around a critical transition point.
In all such cases we emphasize the importance of characterizing the sensitivity not only with respect to system size, but also with respect to protocol time and rate of sweep.
This motivates further research on the scaling laws governing hysteresis behavior beyond adiabatic conditions, including the hysteresis dynamics in bistable systems \cite{Chen2026}.

\bibliography{my_bibliography,supplement_bib}

\clearpage

\section*{End Matter}

\emph{Derivation of master equation}.---The full master equation for the density matrix $\hat{R}$ of the
$N$ three-level atoms is given by
\begin{align}
\partial_{t}\hat{R} & =-i\left[-\delta\hat{J}_{rr}+\Omega(\hat{J}_{rg}+\hat{J}_{gr}),\hat{R}\right]\label{eq:endmatter:ME}\\
 & +\gamma\mathcal{D}_{\hat{J}_{ge}}[\hat{R}]+\gamma_{r}\mathcal{D}_{\hat{J}_{er}}[\hat{R}],\nonumber
\end{align}
with the atomic collective operators given by $\hat{J}_{\mu\nu}=\sum_{n=1}^{N}\left|\mu_{n}\right\rangle \left\langle \nu_{n}\right|$
for states $\mu,\nu\in\{g,e,r\}$. We have already adiabatically eliminated
here the photonic modes (fast decaying with rates $\kappa$ and $\kappa_{r}$), leaving two distinct dissipative channels
operating at rates $\gamma=C\gamma_{eg}$ and $\gamma_{r}=C_{r}\gamma_{re}$,
and our goal is to also eliminate the third level $|r\rangle$ and arrive
at Eq. (\ref{eq:effective master equation}).

The procedure of this adiabatic elimination goes as follows (see full details in the Supplementary Material \cite{supplement_arxiv}). We write an effective coupling in the space of the $\{g,e\}$
states between the state $\left|m\right\rangle \equiv\left|(N/2-m)_{g},(N/2+m)_{e},0_{r}\right\rangle$
and the state $\left|m+1\right\rangle $ with a single excitation
transferred to the $|e\rangle$ state. To first order in $\Omega$ this is
done through the intermediate state $\left|m^{+}\right\rangle \equiv\left|(N/2-m-1)_{g},(N/2+m)_{e},1_{r}\right\rangle$,
and we can deduce the probability amplitude for the whole transition
with analogy to the case of optical pumping on a single 3-level atom.
Specifically, The transition happens with a Rabi drive $\Omega_{m}=\Omega\sqrt{N/2-m}$,
a detuning $\delta_{m}=\delta$ and $r$-level decay $\Gamma_{m}=\gamma_{r}(N/2+m+1)$,
leading to a jump matrix element of

\begin{align}
\left\langle m+1\right|\hat{J}_{\text{pump}}\left|m\right\rangle  & =\frac{\Omega_{m}}{\delta_{m}+i\Gamma_{m}/2}\times\sqrt{\Gamma_{m}}\label{eq:endmatter:effective jump}\\
 & =\frac{\Omega\sqrt{\gamma_{r}(N/2-m)(N/2+m+1)}}{\delta+i\gamma_{r}(N/2+m+1)/2},\nonumber
\end{align}

where $\hat{J}_{\text{pump}}$ denotes the jump operator associated
with pumping. Considering a far detuned laser, $\delta\gg N\gamma_{r}$, this
expression simplifies to

\begin{equation}
\left\langle m+1\right|\hat{J}_{\text{pump}}\left|m\right\rangle \approx\sqrt{w}\left\langle m+1\right|\hat{J}_{+}\left|m\right\rangle ,\label{eq:endmatter:simplified jump}
\end{equation}

in agreement with the jump operator term from Eq. (\ref{eq:effective master equation}), with $w=\gamma_{r}(\Omega/\delta)^{2}$. To summarize, the conditions needed to arrive at (\ref{eq:effective master equation})
are as follows: (i) adiabatic elimination of cavities, requiring $\kappa,\kappa_{r}\gg N\gamma$ in most general scenarios, or $\kappa,\kappa_{r}\gg \sqrt{r \gamma}$ for our metrology protocol at normalized scanning rate $r$ \cite{supplement_arxiv}, (ii) negligible single-atom decays, $C,C_{r}\gg1$, (iii) adiabatic elimination
of $|r\rangle$-excited states, $\delta\gg\sqrt{N}\Omega$, and (iv) the critical regime $w=\gamma_{r}(\Omega/\delta)^{2}\sim\gamma$ (which together
with the previous condition implies $\gamma_{r}\gg N\gamma$).

\emph{Measured observables}.---In the main text we proposed continuously measuring our system in order
to get an enhancement in the sensitivity. This is achievable in our
realization by monitoring either off-axis or on-axis cavity photons.
Each method has its advantages, which we discuss in detail in \cite{supplement_arxiv}
and comment on shortly here.

Since the off-axis emission is effectively uncorrelated between different atoms,
the average off-axis photon-count rate is proportional to the atomic excitation
number $\hat{N}_{e}=\hat{J}_{z}+N/2$. In principle this count is very small, since by construction our model is assumed to have a high
cooperativity $C$ and negligible off-axis emission. To assess the measurement of such a signal, we must then consider the contribution of the resulting shot noise to the
steady-state error in Eq. (\ref{eq: Steady state sensitivity}), finding the total error (see \cite{supplement_arxiv} for details),
\begin{equation}
\Delta w(\alpha)=\Delta w_{0}\times\left(1+O(1/\alpha N)\right),\label{eq:endmatter:shot noise}
\end{equation}
with $\alpha$ denoting the total efficiency of the detector. While at
moderate $N$ this may have a significant effect, we note
that as $N$ gets larger the effect of shot noise becomes negligible.
Intuitively, the reason is that shot noise affects the measured signal
but does not directly shift the transition point, meaning that as
long as enough photons are detected to distinguish between the collective ground
and inverted states the sensitivity is not seriously affected.

From a practical standpoint, however, it might be more attractive to use
the ample information provided by the strong on-axis emission. The
correlated emission from the cavity scales as $\langle\hat{J}_{+}\hat{J}_{-}\rangle\approx(N/2)^{2}-\langle\hat{J}_{z}\rangle^{2}$,
and as a function of pumping, we find that it displays a sharp superradiant peak
at the transition point $w=\gamma$ \cite{supplement_arxiv}. Experimentally this peak is easy to detect, but conceptually one must be careful to apply correct parameter
estimation techniques to extract the sensitivity in this scenario,
since the standard error propagation formula does not apply around
a nonlinear working point. In \cite{supplement_arxiv}
we plot the average cavity emission and define an appropriate
estimator based on averaging symmetrically around the peak. Since
this issue is specific to our setup, we opted to focus on $\hat{J}_{z}$ measurements in the main text.

\emph{Metrology}.---We describe here a method for measuring small deviations in cavity length, stating
the required limits and expected length precision. We envision fixing
the atoms in our system periodically along the cavity mode at positions
a quarter-wavelength away from the mode maxima, such that they lie
in regions of maximum slope of the mode shape. Letting one of the
cavity mirrors to move freely then produces a shift in the atoms'
coupling to the cavity, which translates to a shift in cooperativity
$C$ and hence a shift in the Purcell-enhanced emission $\gamma=C\gamma_{eg}$.
Variations in the critical pumping $w=\gamma$ are then directly connected
to variations in cavity length, and a detailed calculation of absolute length precision gives
\begin{equation}
\Delta L\sim\lambda/N,\label{eq:endmatter:Length precision}
\end{equation}
where $\lambda=2\pi c/\omega_{eg}$ is the transition wavelength.
This is valid as long as the measured length deviations are
smaller than $\lambda$, or else the perturbation shifts the atoms
out of the linear region of the cavity mode shape. Assuming this is
satisfied, measurements of a large atomic ensembles might conceivably
lead to either length or force measurements.

Considering the possible application of measuring
the intrinsic atomic lifetimes of atoms, mentioned in the main text, we provide more details in \cite{supplement_arxiv}.

\clearpage
\onecolumngrid
\setcounter{secnumdepth}{1}

\begin{center}
    \Large\bfseries
    Supplemental Material
\end{center}

\section{Derivation of master equation (1)}

Beginning with the master equation (2) in the main text, the derivation
of the master equation (1) of collectively pumped superradiance must
involve the adiabatic elimination of all the states that
include atoms excited to the state $\left|r\right\rangle $. In the
following, we first present the full derivation, based on a generalized
formulation of adiabatic elimination of open systems (subsection A)
and then consider an alternative, intuitive approach (subsection B). 

\subsection*{A.1. Generalized adiabatic elimination method}

In order to derive the master equation presented in the main text's Eq. (1),
we used an adiabatic elimination scheme first described by \cite{Reiter2012}.
We will present it in a form suited for use in our case.

We start with a system comprised of two sub-spaces, a ground Hilbert
space $\mathcal{H}_{G}$ and an excited Hilbert space $\mathcal{H}_{E}$.
We want to look at a situation where the state is with high probability
in $\mathcal{H}_{G}$, and we weakly excite it into the fast evolving
$\mathcal{H}_{E}$. We include a strong dissipation channel from the
excited to the ground manifold such that all excitations decay fast
to the ground and an effective description inside the ground manifold
only is possible. We also introduce a slight modification to the formalism
by adding a weak dissipation channel which doesn't mix between the
two sub-spaces (this is relevant to our case, where collective decay
from $e$ to $g$ is assumed). Assuming Markovian dynamics, the master
equation for the state $\hat{R}$ (defined on the total Hilbert
space) can be expressed in Lindbald form as
\begin{align}
\dot{\hat{R}} & =-i\left[\hat{H}+\hat{V},\hat{R}\right]+\mathcal{D}_{\hat{K}}[\hat{R}]+\mathcal{D}_{\hat{L}}[\hat{R}],\label{eq:General florentine ME}
\end{align}
where $\hat{H}$ is the Hamiltonian for the excited states, $\hat{V}$
is the exciting perturbation, $\hat{K}$ is the jump operator from
the excited to the ground manifold and $\hat{L}$ is the in-manifold
jump. The formalism can be generalized straightforwardly by adding
modifications such as slow ground evolution, a time dependent perturbation
or multiple dissipation channels. However, for our purposes the formalism
presented is enough. Our goal is to write the out-of-ground-manifold
components of the density matrix in terms of the ground component,
and to write an effective equation for the ground component. To this
end we decompose the density matrix, and likewise the different operators
defining the state evolution, in a block form:
\begin{equation}
\hat{R}=\begin{pmatrix}\hat{\rho}_{EE} & \hat{\rho}_{EG}\\
\hat{\rho}_{GE} & \hat{\rho}_{GG}
\end{pmatrix},\label{eq:decomposition into ground and excited subspaces}
\end{equation}
\begin{equation}
\hat{H}=\begin{pmatrix}\hat{H}_{EE} & 0\\
0 & 0
\end{pmatrix}\ ,\ \hat{V}=\begin{pmatrix}0 & \hat{V}_{EG}\\
\hat{V}_{GE} & 0
\end{pmatrix}\ ,\ \hat{K}=\begin{pmatrix}0 & 0\\
\hat{K}_{GE} & 0
\end{pmatrix}\ ,\ \hat{L}=\begin{pmatrix}\hat{L}_{EE} & 0\\
0 & \hat{L}_{GG}
\end{pmatrix}.
\end{equation}
Following \cite{Reiter2012}, we can then write an effective evolution
equation for the ground components
\begin{align}
\dot{\hat{\rho}}_{GG} & =-i\left[\hat{H}_{\text{eff}},\hat{\rho}_{GG}\right]+\mathcal{D}_{L_{GG}}[\hat{\rho}_{GG}]+\mathcal{D}_{\hat{K}_{\text{eff}}}[\hat{\rho}_{GG}],\label{eq:Effective master equation, general formalism}
\end{align}
where
\begin{align}
\hat{H}_{\text{eff}} & =-\hat{V}_{GE}\frac{\hat{H}_{\text{nh}}^{-1}+\left(\hat{H}_{\text{nh}}^{-1}\right)^{\dagger}}{2}\hat{V}_{EG},\\
\hat{K}_{\text{eff}} & =-\hat{K}_{GE}\hat{H}_{\text{nh}}^{-1}\hat{V}_{EG},
\end{align}
and where we define a non-Hermitian Hamiltonian $\hat{H}_{\text{nh}}$,
which controls the dissipative evolution in the excited manifold
\begin{equation}
\hat{H}_{\text{nh}}=\hat{H}_{EE}-\tfrac{i}{2}\hat{K}_{EG}\hat{K}_{GE}.\label{eq:Nonlinear Hamiltonian}
\end{equation}
Eq. (\ref{eq:Effective master equation, general formalism}) is a valid approximation as long as the
excited manifold is nearly depopulated. This holds when $\left\Vert \hat{H}_{\text{nh}}\right\Vert \gg\left\Vert \hat{V}_{GE}\right\Vert $
(where $\left\Vert \cdot\right\Vert $ represents a typical scale associated with the matrix). We also note that our expression
is slightly modified with respect to the one appearing in \cite{Reiter2012},
due to the fact that we include in-manifold dissipation. This does
not affect the pumping process as long as the rates given by $\hat{L}$
are much lower than the rate of evolution in the excited manifold
(this is enforced if $\left\Vert \hat{H}_{\text{nh}}\right\Vert \gg\left\Vert \hat{L}_{EE}\right\Vert ,\left\Vert \hat{L}_{GG}\right\Vert $
).

\subsection*{A.2. Application to the case of $N$ three level atoms}

Working with the atomic system as described in the paper (with mode
coupling as described above), we can describe the dynamics of the
atom using Eq. (11) in the main text:
\begin{equation}
\dot{\hat{R}}=-i\left[-\delta\hat{J}_{rr}+\Omega\left(\hat{J}_{rg}+\hat{J}_{gr}\right),\hat{R}\right]+\gamma\mathcal{D}_{\hat{J}_{ge}}[\hat{R}]+\gamma_{r}\mathcal{D}_{\hat{J}_{er}}[\hat{R}].\label{eq:ME from paper}
\end{equation}
Physically the state is mostly in the Dicke sector made up of $e$
and $g$ states, while a small percentage (of $O\left(N\Omega^{2}/\delta^{2}\right)$)
have a single atom excited to the $r$ level (there may be further
excitations, but their occupation is of $O\left(\left(N\Omega^{2}/\delta^{2}\right)^{2}\right)$
and can therefore be neglected). This motivates us to define ground
and excited manifolds as above:
\begin{align}
\mathcal{H}_{G} & =\textnormal{span}\left\{ \left|m\right\rangle \equiv\left|\tfrac{N}{2},m\right\rangle _{eg}\right\} _{m=-N/2}^{N/2},\\
\mathcal{H}_{E} & =\hat{J}_{rg}\mathcal{H}_{G}\nonumber \\
 & =\textnormal{span}\left\{ \left|m^{+}\right\rangle \equiv\tfrac{1}{\sqrt{N/2-m}}\hat{J}_{rg}\left|\tfrac{N}{2},m\right\rangle _{eg}\right\} _{m=-N/2}^{N/2-1}.\nonumber 
\end{align}
It is now straightforward to calculate the matrix elements of the various
operators in Eq. (\ref{eq:ME from paper}) needed to apply the formalism
\begin{align}
\left\langle m^{+}\right|\hat{H}_{EE}\left|n^{+}\right\rangle  & =\left\langle m^{+}\right|-\delta\hat{J}_{rr}\left|n^{+}\right\rangle =-\delta\cdot\delta_{mn},\\
\left\langle m^{+}\right|\hat{V}_{EG}\left|n\right\rangle  & =\left\langle m^{+}\right|\Omega\hat{J}_{rg}\left|n\right\rangle =\Omega\sqrt{\tfrac{N}{2}-m}\cdot\delta_{mn},\nonumber \\
\left\langle m\right|\hat{L}_{GG}\left|n\right\rangle  & =\left\langle m\right|\sqrt{\gamma}\hat{J}_{ge}\left|n\right\rangle =\delta_{m+1,n}\sqrt{\gamma\left(\tfrac{N}{2}-n+1\right)\left(\tfrac{N}{2}+n\right)},\nonumber \\
\left\langle m\right|\hat{K}_{GE}\left|n^{+}\right\rangle  & =\left\langle m\right|\sqrt{\gamma_{r}}\hat{J}_{er}\left|n^{+}\right\rangle =\delta_{m-1,n}\sqrt{\gamma_{r}\left(\tfrac{N}{2}+n+1\right)},\nonumber 
\end{align}
and similarly straightforward to calculate the effective Hamiltonian
and jump operator resulting from the adiabatic elimination process
\begin{align}
\left\langle m\right|\hat{K}_{\text{eff}}\left|n^{+}\right\rangle  & =-\left\langle m\right|\hat{K}_{GE}\hat{H}_{\text{nh}}^{-1}\hat{V}_{EG}\left|n^{+}\right\rangle =\frac{\Omega}{\delta}\frac{\sqrt{\gamma_{r}\left(\tfrac{N}{2}-n\right)\left(\tfrac{N}{2}+n+1\right)}}{1+\frac{i\gamma_{r}}{2\delta}\left(\tfrac{N}{2}+n+1\right)}\delta_{m-1,n},\\
\left\langle m\right|\hat{H}_{\text{eff}}\left|n\right\rangle  & =-\left\langle m\right|\hat{V}_{GE}\frac{\hat{H}_{\text{nh}}^{-1}+\left(\hat{H}_{\text{nh}}^{-1}\right)^{\dagger}}{2}\hat{V}_{EG}\left|n\right\rangle =\frac{\Omega^{2}}{\delta}\cdot\frac{\tfrac{N}{2}-m}{1+\frac{\gamma_{r}^{2}}{4\delta^{2}}\left(\tfrac{N}{2}+m+1\right)^{2}}\delta_{mn}.
\end{align}
In the high detuning limit $\delta\gg N\gamma_{r}$ the jump term
simplifies to $\hat{K}_{\text{eff}}=\tfrac{\Omega}{\delta}\sqrt{\gamma_{r}}\hat{J}_{+}$.
The additional energy shift $\hat{H}_{\text{eff}}$ has no observable
effect on the dynamics of population observables (this is proved in
Sec. \ref{sec:steady state}), and so we can safely ignore it. The resulting effective
evolution is given by
\begin{equation}
\dot{\hat{\rho}}=\mathcal{L}[\hat{\rho}]=\gamma\mathcal{D}_{\hat{J}_{-}}[\hat{\rho}]+w\mathcal{D}_{\hat{J}_{+}}[\hat{\rho}],\label{eq:effective equation}
\end{equation}
with $w=\gamma_{r}\left(\Omega/\delta\right)^{2}$, as in the main text.

\subsection*{B. Intuitive approach}
\label{sec:intuitive_approach}

\begin{figure}[t!]
  \centering
\includegraphics[width=0.5\columnwidth]{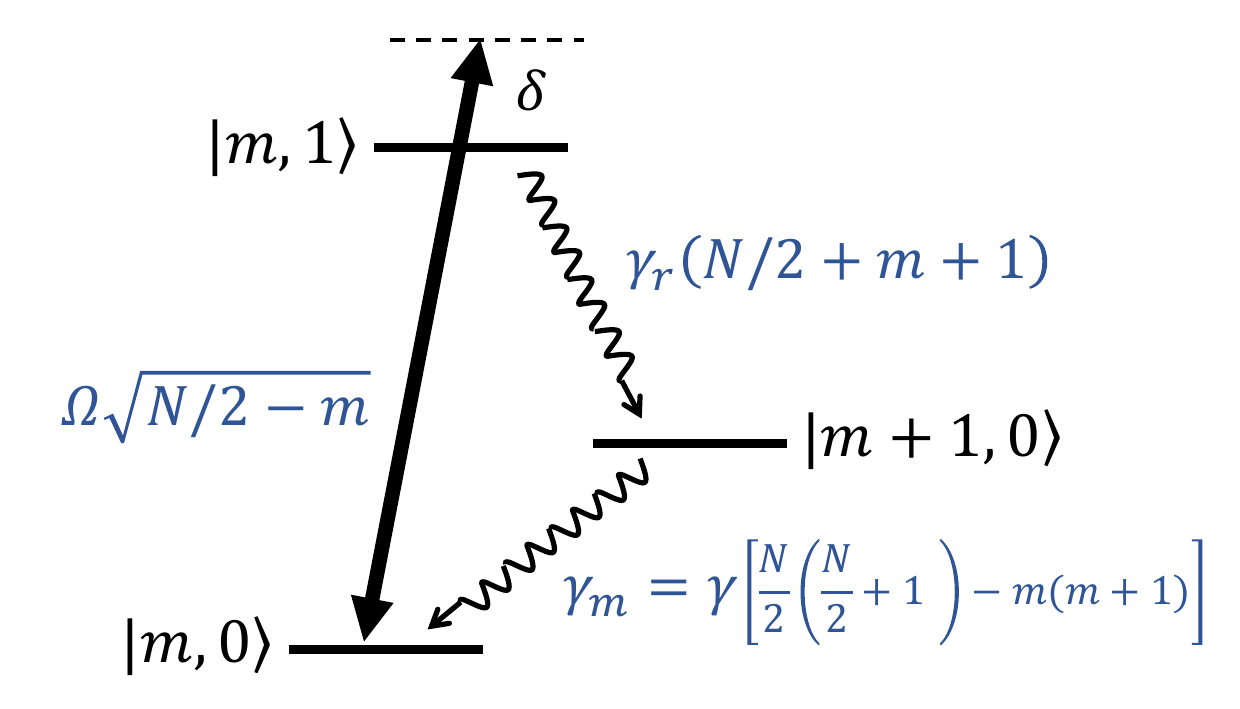}
  \caption{\label{fig: Level scheme} Relevant level scheme considered in Sec. \ref{sec:intuitive_approach}}
\end{figure}

Starting from Eq. (11) of the main text, the essence of the derivation
is to obtain an incoherent term that couples the state $\left|m\right\rangle $
to the state $\left|m+1\right\rangle $ in the subspace of \{$g$,
$e$\} states. To this end, consider the state $\left|m,0\right\rangle \equiv\left|m\right\rangle $,
and the minimal process that can couple it to $\left|m+1,0\right\rangle \equiv\left|m+1\right\rangle $ (in a perturbation theory sense, for weak drive
$\Omega$).
First, the term $\Omega\hat{J}_{rg}$ coupled $\left|m\right\rangle$ to the state
$\left|m,1\right\rangle\equiv\left|m^{+}\right\rangle\propto\hat{J}_{rg}\left|m\right\rangle $, and then
the latter state decays to $\left|m+1,0\right\rangle $ via the collective
dissipation term $\gamma_{r}\mathcal{D}_{\hat{J}_{er}}[\hat{R}]$.
In addition, $\left|m+1,0\right\rangle $ may decay to $\left|m,0\right\rangle $
via the other collective dissipation term $\gamma\mathcal{D}_{\hat{J}_{ge}}[\hat{R}]$.
The coupling and dissipation that gives rise to this process are summarized
in Fig. \ref{fig: Level scheme}. The different rates marked on the figure are
obtained as follows:
\begin{itemize}
\item Drive $\Omega\sqrt{N/2-m}$: The drive term $\Omega\hat{J}_{rg}$
only 'feels' the atoms in states $\left|g\right\rangle $ and $\left|r\right\rangle $.
Therefore, to first order in $\Omega$ it effectively gives a collective
excitation from $N_{g}=N/2-m$ atoms in state $\left|g\right\rangle $
to a single symmetric excitation at state $\left|r\right\rangle $
(the state $\left|m,1\right\rangle $), and so experiences the known
enhancement of square root of the atom number, which here is $N_{g}$.
\item Decay $\gamma_{r}\left(N/2+m+1\right)$: The collective decay term
$\gamma_{r}\mathcal{D}_{\hat{J}_{er}}[\hat{R}]$ only 'feels' the
atoms in states $\left|e\right\rangle $ and $\left|r\right\rangle $.
So, here it describes the decay from a singly excited atom at $\left|r\right\rangle $
to $N_{e}=N/2+m$ atoms at $\left|e\right\rangle $. This decay is
symmetric, i.e. does not distinguish between atoms, since the original
states $\left|m,0\right\rangle $ and $\left|m+1,0\right\rangle $
are symmetric (Dicke states). This means we effectively have
a decay of a symmetric system of $N_{e}+1$ atoms in total, which is commonly known to exhibit the collective
enhancement of the factor of the number of atoms, hence yielding $\gamma_{r}\left(N_{e}+1\right)$.
\item Decay $\gamma_{m}=\gamma \left[N/2\left(N/2+1\right)-m\left(m+1\right)\right]$:
The usual decay between the two Dicke states $\left|m+1\right\rangle $
and $\left|m\right\rangle $ in the \{$g$, $e$\} subspace. A similar
decay $\gamma_{m-1}$ from $\left|m\right\rangle $ to the state $\left|m-1\right\rangle $
outside of this diagram also exists.
\end{itemize}
We now turn to consider the adiabatic elimination of the state $\left|m,1\right\rangle $
which includes the $r$-manifold. Assuming times $t$ much shorter than
the 'free' dynamics of the \{$g$, $e$\}
subspace (i.e. those governed by the term $\gamma\mathcal{D}_{\hat{J}_{ge}}[\hat{R}]$
that does not involves the states $\left|r\right\rangle $, at rate $\sim\gamma_{m}$),
and solving for times much longer than $1/\delta$, the steady state
population of the state $\left|m,1\right\rangle $ (within our coarse-graining
time $1/\delta\ll t\ll1/\gamma_{m}$), has the usual Lorentzian-like
form:
\begin{equation}
p_{m,1}\approx\frac{\left(N/2-m\right)\Omega^{2}}{\delta^{2}+\gamma_{r}^{2}(N/2+m+1)/4}.\label{eq:pm+1}
\end{equation}
Then, since this population decays at rate $\left(N/2+m+1\right)\gamma_{r}$
to $\left|m,1\right\rangle $, the total rate of the incoherent process
from $\left|m,0\right\rangle $ to $\left|m+1,0\right\rangle $ is
given by the product of this population and its decay, yielding $w_{m}=p_{m,1}\left(N/2+m+1\right)\gamma_{r}$.

This incoherent process is then described by the jump operator $\sqrt{w_{m}}\left|m+1\right\rangle \left\langle m\right|$,
which is only within the \{$g$,$e$\} subspace. Summing over the
incoherent processes from different states $\left|m\right\rangle $,
we then obtain the total jump operator $\sum_{m}\left|m+1\right\rangle \left\langle m\right|\sqrt{w_{m}}$.
As we see below, the latter can take an appealing form if we make
the denominator of $p_{m,1}$ independent of $m$, i.e. assuming $\delta\gg\left(N/2+m+1\right)\gamma_{r}$.
Taken together with the separation of timescales we already assumed
in the adiabatic elimination (coarse graining time), we have the condition:
$\delta\gg\left(N/2+m+1\right)\gamma_{r},\gamma_{m},\Omega\sqrt{N/2-m}$
which to be satisfied for all states $m$ requires: 
\begin{equation}
\delta\gg\sqrt{N}\Omega,N\gamma_{r},N^{2}\gamma,
\end{equation}
With this, we can take $p_{m+1}\approx\left(N/2-m\right)\Omega^{2}/\delta^{2}$
in Eq. (\ref{eq:pm+1}) and obtain the total jump operator 
\begin{equation}
\sum_{m}\left|m+1\right\rangle \left\langle m\right|\sqrt{w_{m}}\approx\sqrt{\gamma_{r}}\frac{\Omega}{\delta}\hat{J}_{+}\equiv\sqrt{w}\hat{J}_{+}.
\end{equation}
This indeed yields the optical pumping term $w\mathcal{D}_{\hat{J}_{+}}[\hat{R}]$
in Eq. (1) of the main text with $w=\gamma_{r}\left(\Omega/\delta\right)^{2}$.

It is notable that the constraint $\delta\gg\sqrt{N}\Omega$ ensuring
low excited population, when expressed in terms of $w$, reads $\gamma_{r}\gg Nw$.
At the critical point $w=\gamma$, this leads to a stringent bound
$\gamma_{r}\gg N\gamma$ on the lifetimes of the transitions considered.
This bound can be satisfied given that the slower rate $\gamma$ relates
to a hyperfine transition. For example, in strontium atoms one can
take the relevant levels to be $g\equiv(5s^{2})1S0$, $e\equiv(5s5p)3P0$
and $r\equiv(5s7s)3S1$, in which case the ratio of decay rates is
$\gamma_{r}/\gamma\sim10^{8}$ \cite{Lu2024,Sansonetti2010}. This is well beyond typical
atom numbers used in cavity QED experiments studying the strontium
clock transition \cite{Norcia2016,Norcia2016a}, and so the constraint is comfortably
satisfied. Furthermore, hyperfine systems could potentially be of
interest for exploring our system, where the slow hyperfine transition
is a useful reference with respect to which fast atomic lifetimes
can be measured to a higher precision (see Sec. \ref{sec: applications}).

\section{Realization and consideration of cavity-modes coupling}

\subsection{Single cavity realization}

In the paper we suggested implementing both collective dissipation
and pumping in an atomic ensemble by using two independent cavity
modes to which all the atoms are coupled in an identical way. While in Fig. 1 of the main text we depict, for conceptual simplicity, the possibility to use two distinct cavities for this end, here we describe how a single cavity can suffice. To see this,
assume that we are working in the Rayleigh range and that many wavelengths
fit within it. Consider two longitudinal cavity modes $\hat{a}$ and $\hat{b}$,
resonant with the $|e\rangle\leftrightarrow|g\rangle$ transition
and the $|r\rangle\leftrightarrow|e\rangle$ transition
respectively. Identical coupling to the mode $\hat{a}$ can be achieved if we place
the atoms periodically along the cavity axis at a lattice distance $\lambda_{a}=2\pi c/\omega_{ge}$
from each other (with $\omega_{ge}$ the relevant transition frequency)
such that the atoms are sitting at the maxima of the mode's field
strength. Meanwhile to attain identical coupling with $\hat{b}$ we
need to place the atoms a distance $\lambda_{b}=2\pi c/\omega_{er}$
from each other, which is generally incompatible with the first restriction.

However, if the first condition is satisfied then there is a sufficiently large region along the atomic 'chain' where the mode shape of $\hat{a}$
overlaps well with the mode shape of $\hat{b}$. Specifically, we assume the atoms are trapped at the antinodes of mode $\hat{a}$ and within a total range of $\lambda_{a}\ll\Delta x\ll2\pi/\left|k_{a}-k_{b}\right|$
which is much smaller than the difference of wavenumbers $k_{a}$
and $k_{b}$ of the two cavity modes (with $k_{i}=2\pi/\lambda_{i}$).
This can be done e.g. with a 1D optical lattice \cite{Bohnet2012}
or tweezer array \cite{Yan2023}, noting that such arrangement
can fit multiple trapping sites and atoms thanks to the length scale
separation $\left|k_{a}-k_{b}\right|\ll2\pi/\lambda_{a}$. Then, within
$\Delta x$, both $k_{a}$ and $k_{b}$ appear almost identical, meaning
that the atoms are approximately placed at the antinodes of both cavity
modes simultaneously, so that they exhibit permutation symmetric coupling
to both cavity modes to an excellent degree. 

To arrive at our starting point, Eq. (2) from the main text, we then assume adiabatic elimination the cavity modes, requiring
their relaxation rates to be faster than the atomic system timescale.
This implies $\kappa,\kappa_{r}\gg N\gamma$ in the most general case,
or $\kappa,\kappa_{r}\gg\gamma$ when operating very close to the
critical point (see Sec. \ref{sec:eliminating_kappa} below). These conditions can generally be satisfied for a cavity
of length $L\lesssim1$cm and finesse $\mathcal{F}\sim10^{5}$,
which is typical of cavity QED setups such as \cite{Thompson1998}.

\subsection{Individual decay and pumping due to off axis emission}

While the collective decay and pumping originates in the joint emission
of atoms via cavity modes, individual decay and pumping processes
also emerge due to emission to off-cavity-axis modes from the two
corresponding transitions at rates $\gamma_{i}=\gamma_{eg}$ and $w_{i}=\gamma_{rg}(\Omega/\delta)^{2}$,
respectively. Recalling the cooperativities of the cavity modes, $C$
and $C_{r}$, these individual processes can also be written as $\gamma_{i}=\gamma/C$
and $w_{i}=w/C_{r}$ with $w$ and $\gamma$ the rates of the cavity
enhanced processes from Eq. (1) in the main text. While in the main
text we assume that the individual processes are much weak, $C,C_{r}\gg1$
and that the experiment ends before their effect is appreciable ($T<\gamma_{eg}^{-1}=\gamma_{i}^{-1}$),
here we address their effect in mode detail. In particular, we include
these individual effects in the master equation (1) and derive
approximate mean-field equations for the average spin and total
spin length, arriving at
\begin{align}
\partial_{t}N_{\text{eff}} & =N\left(w_{i}-\gamma_{i}\right)s_{z}-\frac{w_{i}+\gamma_{i}}{2}\left(1+s_{z}^{2}\right)N_{\text{eff}},\\
\partial_{t}z & =\frac{w-\gamma}{2}N_{\text{eff}}-\left(w+\gamma+\frac{w-\gamma}{2}N_{\text{eff}}z\right)z\nonumber \\
 & +\left(w_{i}-\gamma_{i}\right)\frac{N}{N_{\text{eff}}}-\left(\frac{w_{i}+\gamma_{i}}{2}\left(1-z^{2}\right)+\left(w_{i}-\gamma_{i}\right)\frac{N}{N_{\text{eff}}}z\right)z,\nonumber 
\end{align}
where $N_{\text{eff}}\equiv2\sqrt{\langle\hat{\boldsymbol{J}}^{2}\rangle}$
is the effective number of collectively interacting atoms and $z\equiv\frac{2}{N_{\text{eff}}}\langle\hat{J}_{z}\rangle$
is the normalized spin projection. In our system these are weaker than the
collective rates $w,\gamma$ by a factor $C$ (taking e.g. $C_r=C$), and thus near the transition
($\Delta w=w-\gamma=O\left(\gamma/N\right)$),
\begin{align}
\partial_{t}N_{\text{eff}} & =-\frac{\gamma}{C}\left(1-z^{2}\right)N_{\text{eff}},\\
\partial_{t}z & =\frac{N_{\text{eff}}\Delta w}{2}-\left(2\gamma\left(1+\frac{1-z^{2}}{2C}\right)+\frac{N_{\text{eff}}\Delta w}{2}z\right)z.\nonumber 
\end{align}
For $C\gg1$ the effective atom number decays at a rate $\sim\gamma_i=\gamma/C$
and our ideal transition width is changed to $\Delta w_{0}\sim\gamma/N_{\text{eff}}$.
After a time $T\sim C\gamma^{-1}$ the metrological advantage of our
protocol is lost, and the system should be re-initialized to a fully
excited state. The key figure for optimizing the experiment time is
thus the cavity's cooperativity - we note though that according to the main text's Eq. (10) even for a moderately
high value our protocol still approaches Heisenberg-like sensitivity,
with cavity-related improvements scaling as $O(1/\sqrt{C})$.
This means that improving $C$ has the same effect as repeating the
experiment many times.

An advantage of our non-adiabatic protocol with respect to adiabatic
preparation is that it allows us to look for the transition over a
much wider range of pumping values. This can be seen from the above
analysis, where we concluded that the experiment time is ultimately
limited by the cooperativity $C$. Since our protocol requires us
to have enough time to preform several sweeps, this creates a restriction
$N_{s}\sim\dot{w}T/\Delta w_{\text{scan}}\gg1$ which translates to
a bound on the maximum sweeping range;
\begin{equation}
\frac{\Delta w_{\text{scan}}}{\Delta w_{0}}\lesssim rT=C r/\gamma.
\end{equation}
Under adiabatic conditions ($r\sim\gamma$), finding the transition
to a precision $\Delta w_{0}$ would require knowing in advance where
the transition lies to an accuracy of $\Delta w_{\text{scan}}\sim C\Delta w_{0}$,
making the utility of our protocol very limited. However, our beyond
- adiabatic analysis shows that much larger sweeping ranges are possible,
provided that we sufficiently increase the sweeping rate and correctly
analyze the information given to us by the hysteresis loop. Our protocol
then requires little prior information about the location of the transition, making it a relevant platform for global sensing. 

\subsection{Elimination of the cavity modes}
\label{sec:eliminating_kappa}

We recall that the starting point of our model, Eq. (1) above, is
written within a Markov approximation wherein the cavity is eliminated
by assuming its relaxation rate $\kappa$ to be faster than the typical
time scale of the atoms. The latter is bounded by the superradiant
emission rate $N\gamma$, yielding the validity condition $\kappa\gg N\gamma$
mentioned in the main text. Notably however, this condition corresponds
to a worst-case scenario, since the relaxation rate of the atomic
system during our sweeping protocol is typically much slower than
$N\gamma$. In particular, for a sweeping in an interval $\Delta w=|w-\gamma|\gg\gamma/N$
around the transition, we recall the relaxation rate (Eq. (6) in main
text),
\begin{equation}
\lambda=N\Delta w.
\end{equation}
While this expression is
capable of reaching $O\left(N\gamma\right)$, in our setup
the maximum relaxation will be set by the maximum sweeping amplitude.
Recalling that in our protocol we vary this amplitude to accommodate
the size of the growing hysteresis loop,
\begin{align}
\Delta w_{\text{scan}} & =A\times\max\left\{ \Delta w_{0},w_{+}-w_{-}\right\} ,
\end{align}
(with $A\gtrsim1$) we conclude that a more realistic condition for adiabatic elimination
of the cavities is (using Eq. (7) from the main text for $w_{+}-w_{-}$), 
\begin{equation}
\kappa\gg N\Delta w_{\text{scan}}\sim A\gamma\times\begin{cases}
1 & r\ll\gamma\\
\left(r/\gamma\right)^{\eta} & r\gg\gamma
\end{cases}.
\end{equation}
Recalling that $\eta\approx0.5$, this means that setting $\kappa$ to be sufficiently larger than $\sqrt{\gamma r}$ guarantees the validity of eliminating the cavity from the dynamics. For comparison, \cite{Norcia2016} uses a cavity
with $\kappa\sim1\text{MHz}$, only somewhat larger than the collective
decay rate $N\gamma\sim0.3\text{MHz}$ but larger than the single
atom decay by a factor $\kappa/\gamma\sim10^{5}$, making such a setup relevant in our case.

\section{Analytical calculation of steady state and numerical validation of
its uniqueness}
\label{sec:steady state}

In Eq. (2) of the text we stated that the system has a unique steady state of the form:
\begin{equation}
\hat{\rho}_{\beta}\propto e^{-\beta\hat{J}_{z}},
\end{equation}
with $\beta=-\ln\left(w/\gamma\right)$. We show that this is a solution
by direct substitution. To this end we use the identities:
\begin{equation}
\begin{array}{c}
e^{\beta\hat{J}_{z}}\hat{J}_{-}e^{-\beta\hat{J}_{z}}=e^{-\beta}\hat{J}_{-}\ ,\ e^{\beta\hat{J}_{z}}\hat{J}_{+}e^{-\beta\hat{J}_{z}}=e^{\beta}\hat{J}_{+},\\
\left[\hat{J}_{+}\hat{J}_{-},e^{-\beta\hat{J}_{z}}\right]=\left[\hat{J}_{-}\hat{J}_{+},e^{-\beta\hat{J}_{z}}\right]=0.
\end{array}
\end{equation}
Moving on to the calculation:
\begin{align}
\dot{\hat{\rho}}_{\beta} & \propto\gamma\mathcal{D}_{\hat{J}_{-}}\left[\hat{\rho}_{\beta}\right]+w\mathcal{D}_{\hat{J}_{+}}\left[\hat{\rho}_{\beta}\right]\\
 & =\gamma\left(\hat{J}_{-}e^{-\beta\hat{J}_{z}}\hat{J}_{+}-\tfrac{1}{2}\left\{ \hat{J}_{+}\hat{J}_{-},e^{-\beta\hat{J}_{z}}\right\} \right)+\gamma e^{-\beta}\left(\hat{J}_{+}e^{-\beta\hat{J}_{z}}\hat{J}_{-}-\tfrac{1}{2}\left\{ \hat{J}_{-}\hat{J}_{+},e^{-\beta\hat{J}_{z}}\right\} \right)\nonumber \\
 & =\gamma e^{-\beta\hat{J}_{z}}\left(\left(e^{\beta\hat{J}_{z}}\hat{J}_{-}e^{-\beta\hat{J}_{z}}\right)\hat{J}_{+}-\hat{J}_{+}\hat{J}_{-}\right)+\gamma e^{-\beta}\cdot e^{-\beta\hat{J}_{z}}\left(\left(e^{\beta\hat{J}_{z}}\hat{J}_{+}e^{-\beta\hat{J}_{z}}\right)\hat{J}_{-}-\hat{J}_{-}\hat{J}_{+}\right)\nonumber \\
 & =\gamma e^{-\beta\hat{J}_{z}}\left(e^{-\beta}\hat{J}_{-}\hat{J}_{+}-\hat{J}_{+}\hat{J}_{-}\right)+\gamma e^{-\beta\hat{J}_{z}}\left(e^{-\beta}\cdot e^{\beta}\hat{J}_{+}\hat{J}_{-}-e^{-\beta}\hat{J}_{-}\hat{J}_{+}\right)\nonumber \\
 & =0.\nonumber 
\end{align}
So this is indeed a steady state. To be sure of its uniqueness, consider
first the following decomposition of the space of density matrices
into off-diagonals:
\begin{equation}
\hat{\rho}=\sum_{q=-N}^{N}\hat{\rho}^{\left(q\right)}\ ,\ \hat{\rho}^{\left(q\right)}\in\text{Diag}^{\left(q\right)}=\begin{cases}
\text{span}\left\{ \left|m\right\rangle \left\langle m+q\right|\right\} _{m=-N/2}^{N/2-q} & q\ge0\\
\text{Diag}^{\left(-q\right)\dagger} & q<0
\end{cases}.\label{eq:Decomposition into off diagonals}
\end{equation}
\begin{figure}[t!]
  \centering
\includegraphics[width=0.5\columnwidth]{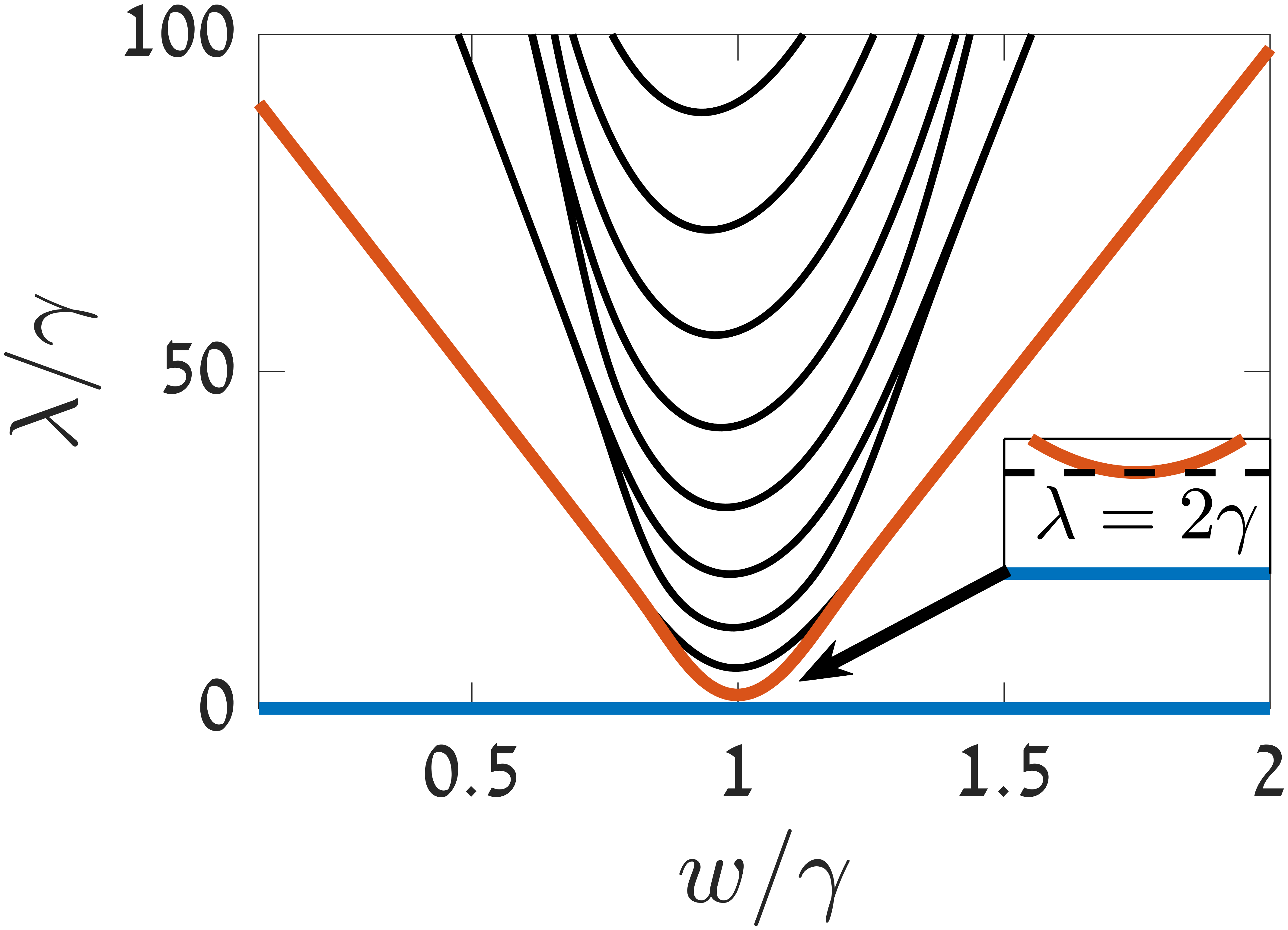}
  \caption{\label{fig: Lindbaldian spectrum} Lindbaldian spectrum of $\mathcal{L}$ in the $\textnormal{Diag}^{\left(0\right)}$
sector, for $N=100$. The blue curve represents the zero mode, the red curve represents
the Lindbaldian gap and the black curves show the higher dissipative
modes. Inside the closeup of the transition point we see a minimal
gap of $\lambda=2\gamma$.}
\end{figure}
One can be easily convinced that the evolution operator $\mathcal{L}\equiv\gamma\mathcal{D}_{\hat{J}_{-}}+w\mathcal{D}_{\hat{J}_{+}}$
maps elements of $\text{Diag}^{\left(q\right)}$ to elements of $\text{Diag}^{\left(q\right)}$,
and thus decomposes as a direct sum over these spaces. This means
that the modes of the system can be found by diagonalizing $\mathcal{L}$
over each subspace individually, which is easily done on a computer.
In Fig. \ref{fig: Lindbaldian spectrum} the resulting spectrum for the diagonal subspace
$\text{Diag}^{\left(0\right)}$ is shown, with only one zero eigenvalue
(representing the steady state) and with the smallest nonzero eigenvalue
having a gap of $\lambda=2\gamma$. The spectrum in all other subspaces
is similarly gapped, with the smallest gap appearing in $\text{Diag}^{\left(\pm1\right)}$ and having a value of $\lambda=\gamma$.
This ensures us that the system has only one steady state, and that
any perturbation decays at most after a time $\tau\sim1/\gamma$.

Another consequence of this decomposition is that we are justified in neglecting the nonlinear light shift introduced by the adiabatic elimination process, as long as we work only in the diagonal sector (meaning that we only study population dynamics, as we do in the main text). This is because its projection on diagonal elements is null:
\begin{equation}
\left\langle m\right|[\hat{\rho},\hat{H}_{\text{eff}}]\left|m\right\rangle \equiv0
\end{equation}

\section{Methods of measuring the transition point}

In the paper we suggested measuring the transition point from off-cavity-axis emission, which might appear to be problematic due to the assumed
weakness of the signal. Here we show that the shot noise resulting
from such weakness does not substantially increase the error in the
protocol, as long as enough atoms are involved.

The contribution of shot noise to the detection of off-axis emission
can be modeled by a Poisson process whose detection rate fluctuates
as $\hat{\Gamma}_{\text{off-axis}}=\gamma_{ge}\hat{N}_{e}$. The statistics
of the detector are then given by 
\begin{align}
\langle N_{D}\rangle & =\alpha\langle N_{e}\rangle\\
\text{var}[N_{D}] & =\alpha^{2}\text{var}[N_{e}]+\alpha\langle N_{e}\rangle,\nonumber 
\end{align}
where $\alpha=\eta\gamma_{eg}T$ is the total detection efficiency,
determined by the photon collection efficiency $\eta$, the single
atom decay rate $\gamma_{eg}$ and the experiment time $T$. Given
that $T\lesssim\frac{1}{\gamma_{eg}}$, $\eta\ll1$ and thus $\alpha\ll1$,
it might appear that off-axis detection is highly susceptible to shot
noise. We note however that the relevant quantity is not the emission,
but rather the critical pumping $w$ at which population inversion
occurs. Since this is determined by locating a sharp transition, errors
in $w$ are mostly unaffected by errors in the population itself,
as long as there's enough signal to differentiate the ground from
the inverted state. 

To see this quantitatively we can calculate $\Delta w$ as a function
of total efficiency $\alpha$:
\begin{equation}
\Delta w(\alpha)=\frac{\sqrt{\text{var}[N_{D}]}}{\left|\partial_{w}\langle N_{D}\rangle\right|}=\Delta w_{0}\times\sqrt{1+\frac{\langle N_{e}\rangle}{\alpha\text{var}[N_{e}]}}.
\end{equation}
Around the critical point we know that $\langle N_{e}\rangle\equiv N/2$
and that $\text{var}[N_{e}]\sim N^{2}$. Thus the ideal error $\Delta w_{0}\sim\gamma/N$
is widened by a factor of $1+O(1/\alpha N)$, which can be mitigated
by increasing the number of atoms.

\begin{figure}[t!]
  \centering
\includegraphics[width=0.5\columnwidth]{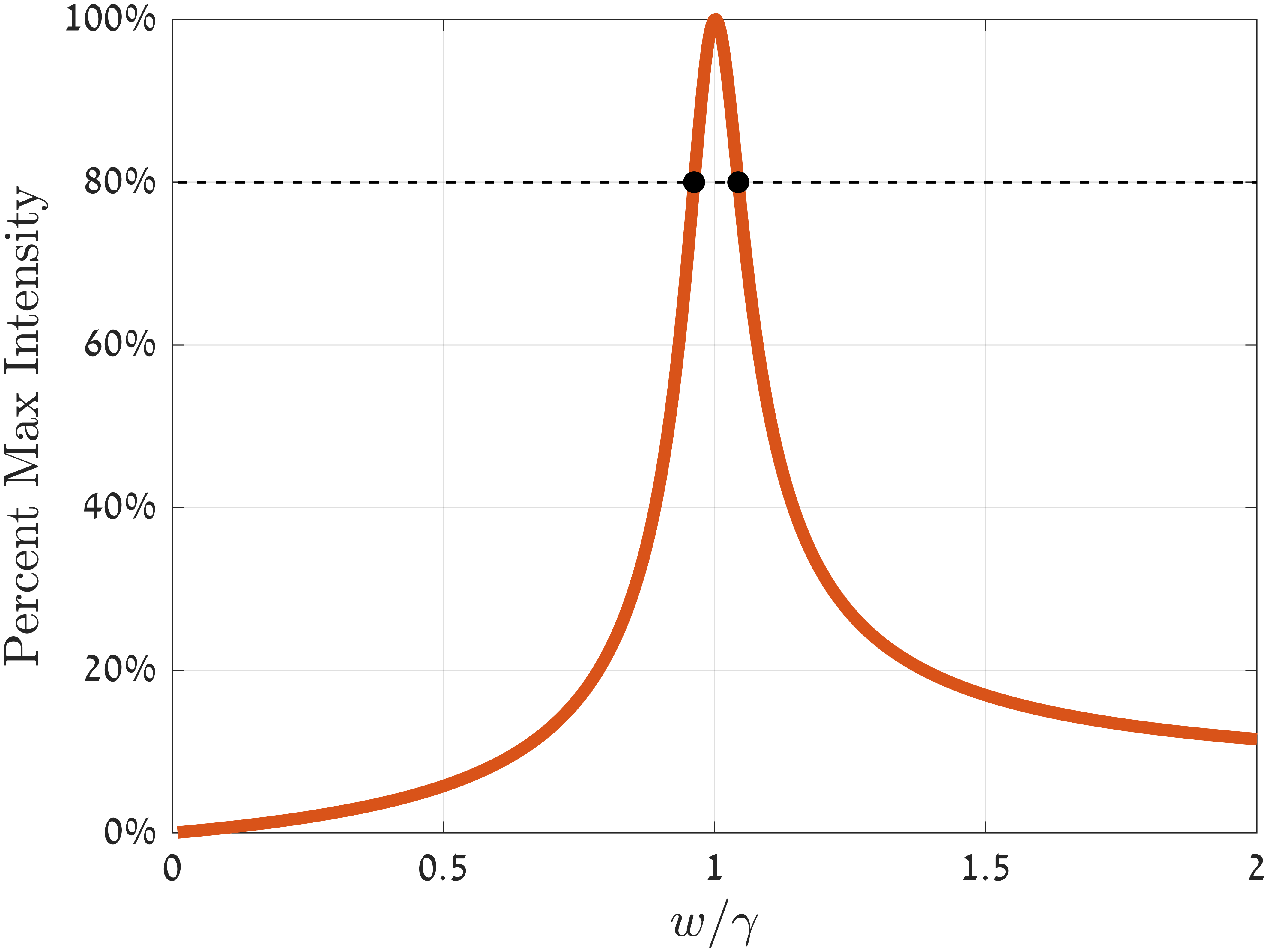}
  \caption{\label{fig: intensity} Average cavity-axis emission rate, normalized to maximum intensity, vs. pumping rate. normalized to the transition point, depicted in the red plot for $N=100$. At the 80\% level (dashed black line) are marked two crossing points (black), which are used to define the estimator for the transition point.}
\end{figure}

If however $\alpha N\ll1$ (meaning that the efficiency is much too
low to detect any off-axis photons), we can still extract the critical
pumping from the overwhelmingly stronger cavity-axis emission. This
signal measures the collective radiation to the cavity mode, related
to excitation number as $\hat{\Gamma}_{\text{cavity-axis}}=\gamma\hat{N}_{e}(N-\hat{N}_{e}+1)$.
In Fig. \ref{fig: intensity} we plot $\langle\hat{\Gamma}_{\text{cavity-axis}}\rangle$
as a function of $w$, showing a superradiant peak in intensity around
the population inversion point. 

An estimator for the critical pumping can be defined by measuring
the two crossing points at a certain intensity threshold (as in Fig. \ref{fig: intensity})
and taking their average. Our model predicts that an ideal sensitivity
is reached when sampling the points at $80{\%}$ of maximum
intensity, and that the resulting uncertainty is $\Delta w'\approx2\Delta w_{0}$.
An analysis of how the uncertainty in this protocol changes due to
hysteresis effects is however beyond the scope of the discussion,
and we leave it for future work. 

\section{Cramér--Rao bound for the sensitivity}

In the text's Eq. (4) we gave an estimate of the steady state sensitivity to changes in the pumping rate, finding it to be of
$O\left(1/N\right)$. We now make this more exact by using the Cramér--Rao
bound with respect to the estimated parameter $\beta=-\ln\left(w/\gamma\right)$.
To this end we note that the distribution of $\hat{J}_{z}$ values
is given by
\begin{align}
p_{m}\left(\beta\right) & =\left\langle m\right|\hat{\rho}_{\beta}\left|m\right\rangle =\frac{1}{Z\left(\beta\right)}e^{-\beta m},\\
Z\left(\beta\right) & =\frac{\sinh\left(\frac{N+1}{2}\beta\right)}{\sinh\left(\frac{1}{2}\beta\right)}.\label{eq:partition function}
\end{align}
Since the state is diagonal in this basis for all $\beta$, we know
that the quantum Fisher information with respect to $\beta$ is the same as the
classical Fisher information \cite{Tan2019}, and we can compute it as
\begin{align}
I_{\beta} & =\sum_{m}p_{m}\left(\beta\right)\left(\partial_{\beta}\ln p_{m}\left(\beta\right)\right)^{2}\\
 & =\sum_{m}p_{m}\left(\beta\right)\left(m+\partial_{\beta}\ln Z\left(\beta\right)\right)^{2}=\text{tr}\left(\hat{\rho}_{\beta}\left(\hat{J}_{z}-\langle\hat{J}_{z}\rangle\right)^{2}\right)\nonumber \\
 & =\text{Var}[\hat{J}_{z}].\label{eq:Fisher information} 
\end{align}
We now have a formal bound on the possible precision with which we
can estimate $\beta$:
\begin{equation}
\Delta\beta\ge\frac{1}{\sqrt{I_{\beta}}}=\frac{1}{\sqrt{\text{Var}[\hat{J}_{z}]}}.
\end{equation}
Lastly, we note that at the transition the variance as presented in
Eq. (\ref{eq:Fisher information}) achieves the value
\begin{equation}
\text{Var}[\hat{J}_{z}]|_{\beta=0}=-\partial_{\beta}\langle\hat{J}_{z}\rangle=\frac{N^{2}}{12},
\end{equation}
so inputting this result (and replacing $\Delta\beta=\frac{\Delta w}{w}$),
we get the bound
\begin{equation}
\frac{\Delta w}{w}\ge\frac{\sqrt{12}}{N}.
\end{equation}
Aditionally, one can show that $\langle\hat{J}_z\rangle$ is a maximum likelihood estimator for $\beta$, ensuring that the bound is saturated by the metrology scheme presented in the main text.  

\section{Metrological applications: Atomic lifetime estimation}
\label{sec: applications}

Our procedure allows us to measure a critical pumping rate $w=\gamma$
to a precision of $O(1/N)$. In terms of intrinsic parameters, the
critical condition reads:
\begin{equation}
\left(\frac{\Omega}{\delta}\right)^{2}=\frac{C\gamma_{eg}}{C_{r}\gamma_{re}},
\end{equation}
where $\Omega$, $\delta$ are the laser's Rabi frequency and detuning
from the $r$ state, $C$,$C_{r}$ are the cooperativities of the
cavity modes $\hat{a}$ and $\hat{b}$ respectively and $\gamma_{eg}$,
$\gamma_{re}$ are the free space decay rates of the atomic transitions
in resonance with said cavity modes.

Depending on which parameters are known, we can extract information
on the others. Assuming the ratio $\Omega/\delta$ is known (either
by fine calibration or by measuring the population of $r$-excited
states), and that the ratio $C/C_{r}$ is determined by other means,
we can then measure the ratio $\gamma_{eg}/\gamma_{re}$ to a good
precision. If $\gamma_{eg}$ represents a slow rate which is easier
to measure directly, then we can use it as a reference with respect
to which much faster decay rates (represented by $\gamma_{re}$) can
be measured. Given the relatively modest precision standards for current lifetime measurements \cite{Tanner1994,Traebert2024,Monier2010} (of order $\gtrsim0.1\%$ for elements such as strontium \cite{Sansonetti2010}) and the potential application of lifetime data to areas such as astrophysical measurements \cite{Pickering2011}, cold atom experiments and tests of CP violations \cite{Nemouchi2009}, this measurement
could then be of scientific interest.

To give a quantitative estimate for the potential improvement in precision,
Let us take for example \ce{^87Sr} atoms (as in \cite{Norcia2016,Norcia2016a})
and choose our target levels to be $g\equiv(5s^{2})1S0$, $e\equiv(5s5p)3P0$
and $r\equiv(5s7s)3S1$. $\gamma_{eg}\sim10\text{mHz}$ is then a
hyperfine transition, and $\gamma_{re}\sim1\text{MHz}$ \cite{Sansonetti2010}.
By propagation of error, we can estimate the achievable relative error
in $\gamma_{re}$ in terms of that of $\gamma_{eg}$ and the atom
number:
\begin{equation}
\Delta\gamma_{re}/\gamma_{re}\approx\sqrt{\left(\Delta\gamma_{eg}/\gamma_{eg}\right)^{2}+\frac{12}{N^{2}}}
\end{equation}
Current estimates on \ce{^{87}Sr}'s hyperfine transition seems to
stand at $\Delta\gamma_{eg}/\gamma_{eg}\sim0.1\%$ \cite{Lu2024}, while
for the fast transition $\Delta\gamma_{re}/\gamma_{re}\sim10\%$ \cite{Sansonetti2010}.
It is therefore in principle possible to improve current estimates
by an order of magnitude or two. Furthermore, given our procedure
improvements in the measurement of hyperfine decay rates can have
a direct effect on the precision of faster ones, potentially paving
the way for a better understanding of atomic dissipation mechanisms.

\section{Calculation of the relaxation rate $\lambda$ by numerical diagonalization
and by cumulant methods}

Since calculation of the mean inversion $\langle\hat{J}_{z}\left(t\right)\rangle=\text{tr}\left(\hat{J}_{z}\hat{\rho}\left(t\right)\right)$
requires only diagonal matrix elements (which as stated in Sec. (\ref{sec:steady state}) evolve independently in the $\text{Diag}^{\left(0\right)}$ subspace),
the spectrum shown in Fig. \ref{fig: Lindbaldian spectrum} contains all rates governing
the evolution of the mean, and in particular the lowest nonzero eigenvalue
$\lambda\left(w\right)$ represents the exact asymptotic rate at which
$\langle\hat{J}_{z}\left(t\right)\rangle$ decays. We contrast this
with an approximate analytical expression which we obtain by analyzing
the two time correlation $C_{zz}$ given by
\begin{align}
C_{zz}\left(\tau\right) & \equiv\langle\hat{J}_{z}\left(t+\tau\right)\hat{J}_{z}\left(t\right)\rangle-\langle\hat{J}_{z}\left(t+\tau\right)\rangle\langle\hat{J}_{z}\left(t\right)\rangle.
\end{align}
At long times these correlations decay at a rate $C_{zz}\left(\tau\right)\sim e^{-\lambda\left(w\right)\tau}$,
which can be approximately calculated by writing the rate equation
for $C_{zz}$ obtained from the master equation in terms of higher order correlations
\begin{equation}
\partial_{\tau}C_{zz}\left(\tau\right)=\left(w-\gamma\right)\left(\langle\hat{J}_{z}^{2}\left(t+\tau\right)\hat{J}_{z}\left(t\right)\rangle-\langle\hat{J}_{z}^{2}\left(t+\tau\right)\rangle\langle\hat{J}_{z}\left(t\right)\rangle\right)-\left(w+\gamma\right)C_{zz}\left(\tau\right),
\end{equation}
and then breaking these correlations into lower order ones by neglecting
third order cumulants \cite{Kubo1962}
\begin{align}
\langle\hat{J}_{z}^{2}\left(t+\tau\right)\hat{J}_{z}\left(t\right)\rangle & \approx2\langle\hat{J}_{z}\left(t+\tau\right)\rangle C_{zz}\left(\tau\right)+\langle\hat{J}_{z}^{2}\left(t+\tau\right)\rangle\langle\hat{J}_{z}\left(t\right)\rangle.
\end{align}
In this way we reach an expression for the decay rate
\begin{align}
\partial_{\tau}C_{zz} & =-2\left(\tfrac{w+\gamma}{2}+\left(w-\gamma\right)\langle\hat{J}_{z}\rangle\right)C_{zz}\rightarrow\\
\lambda\left(w\right) & =w+\gamma+2\left(w-\gamma\right)\langle\hat{J}_{z}\rangle\left(w\right),
\label{eq: relaxation}\end{align}
where the mean inversion $\langle\hat{J}_{z}\rangle$
can be calculated directly by using the partition function in Eq. (\ref{eq:partition function}):
\begin{align}
\langle\hat{J}_{z}\rangle & =-\partial_{\beta}\ln Z\left(\beta\right)|_{\beta=-\ln\left(w/\gamma\right)}\\
 & =\frac{N+1}{2}\frac{(w/\gamma)^{N+1}+1}{(w/\gamma)^{N+1}-1}-\frac{1}{2}\frac{w/\gamma+1}{w/\gamma-1},\nonumber 
\end{align}
and in the limit $N\gg1$, $\left|w/\gamma-1\right|\ll1$ is given by
\begin{equation}
\langle\hat{J}_{z}\rangle\rightarrow\frac{N/2}{\tanh\left(N\left(w/\gamma-1\right)/2\right)}-\frac{1}{w/\gamma-1}.
\end{equation}
This expression for $\lambda\left(w\right)$ was compared in Fig. 2 of the paper against the lowest nonzero eigenvalue
appearing in Fig. \ref{fig: Lindbaldian spectrum}. While the numerical and analytical
curves approximately agree, we note that there is a discrepancy between
them of order $\gamma$. As $N\rightarrow\infty$ this discrepancy
persists, but the scale of $\lambda$ away from the transition grows
with $N$ so as to make this error negligible. At the transition this
error cannot be said to be negligible, because the relaxation rate
approaches a minimum value of $\lambda=2\gamma$. However, both curves
agree on this minimum value such that the asymptotic expression in Eq. (\ref{eq: relaxation}) is correct.

\section{Independence of hysteresis shape scaling on atom number}

\begin{figure}[t!]
  \centering
\includegraphics[width=0.5\columnwidth]{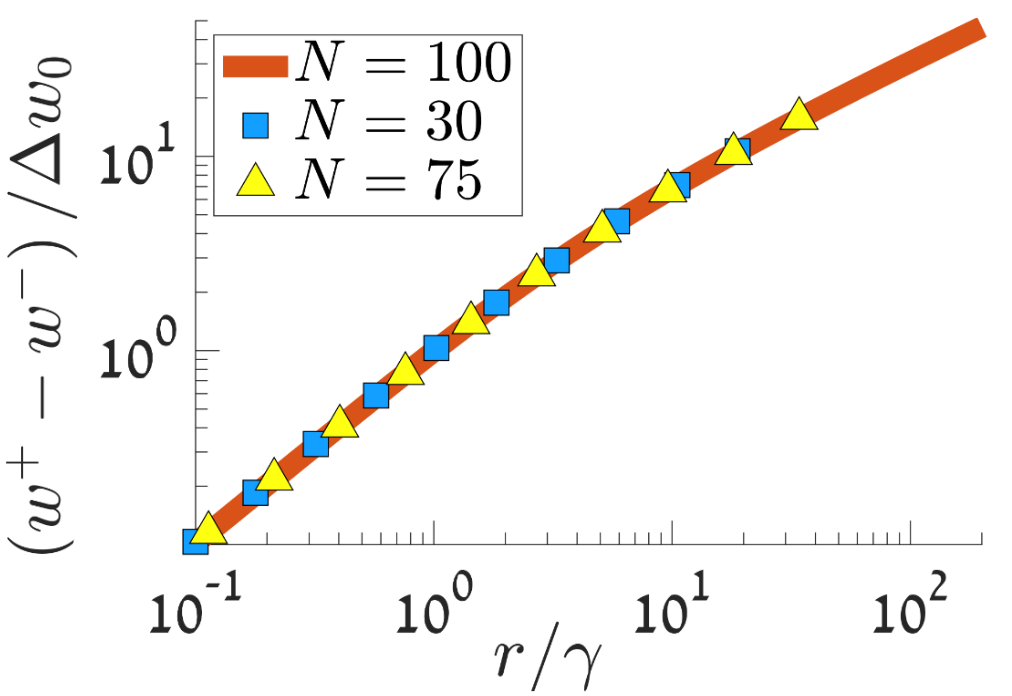}
  \caption{\label{fig: Ndependence} The different curves of normalized hysteresis width obtained for different
$N$ values are seen to collapse on each other in accordance with
the prediction of the main text's Eq. (7).}
\end{figure}

An important property of the scaling laws which we postulated in Eq.
(7) of the main text is their independence of atom number, relative
to the adiabatic sensitivity $\Delta w_{0}$. This is something to
be validated by comparing the $r$ dependence of $(w^{+}-w^{-})/\Delta w_{0}$
and $\Delta w^{\pm}/\Delta w_{0}$ for different atom numbers $N$,
which we have done. The results are indeed independent of atom
number for moderately high $N$, as can be seen in Fig. \ref{fig: Ndependence} where we show the hysteresis width
plot for $N=30,75,100$.

\section{Asymptotic symmetry of the hysteresis loop}

To demonstrate the asymptotic symmetry property mentioned in the paper
we start from the time dependent master equation for the atomic population
$p_{m}\equiv\left\langle m|\hat{\rho}|m\right\rangle $:

\begin{align}
\partial_{t}p_{m} & =\Gamma_{m+1}p_{m+1}-\Gamma_{m}p_{m}+w(t)/\gamma\cdot\left(\Gamma_{m}p_{m-1}-\Gamma_{m+1}p_{m}\right),\\
\Gamma_{m} & \equiv\gamma\left(\tfrac{N}{2}+m\right)\left(\tfrac{N}{2}-m+1\right),\nonumber \\
w(t) & \equiv\gamma+\Delta w_{0}rt=\gamma\left(1+\tfrac{\sqrt{12}}{N}rt\right).\nonumber 
\end{align}

This discrete master equation has a continuum limit defined by taking
$x=\frac{2m}{N}$ to be a continuous parameter in the range $[-1,1]$
(with increment $\tfrac{2}{N}\rightarrow dx$), and expressing $p_{m}$
as a continuous distribution on $x$:
\begin{align}
p_{m} & \overset{N\rightarrow\infty}{\longrightarrow}f(x)dx.
\end{align}
Under this limit it is elementary to show that our master equation
approaches the following Fokker-Planck equation:
\begin{equation}
\tfrac{1}{\gamma}\partial_{t}f=\partial_{x}\left(1-x^{2}\right)\partial_{x}f-\sqrt{3}rt\cdot\partial_{x}\left(1-x^{2}\right)f.
\end{equation}
This equation is manifestly invariant under a simultaneous inversion
of population ($x\rightarrow-x$) and direction of sweep ($r\rightarrow-r$),
with the implication that the backward-sweep solution is the inverted
image of of the forward one.

\section{Hysteresis width power law}

In the main text we presented a scaling law for the hysteresis width as
a function of the scanning rate $r$, as presented in Eq. (8) of the
paper. In general, if the gap of a critical system scales as $\lambda_{0}\cdot\left(\Delta w/\Delta w_{0}\right)^a$,
with $a$ an exponent determined by the system's dynamics, then
at the boundary of hysteresis 
\begin{align}
\text{Scanning rate} & =\frac{\Delta w_{0}\cdot r}{\Delta w_{H}}\sim\lambda_{0}\cdot\left(\Delta w_{H}/\Delta w_{0}\right)^a=\text{Relaxation rate}\\
\rightarrow\Delta w_{H} & \sim\Delta w_{0}\cdot\left(r/\lambda_{0}\right)^{\eta}\text{ with }\text{\ensuremath{\eta=}\ensuremath{\frac{1}{1+a}}},\nonumber 
\end{align}
where $\Delta w_{H}=w_{+}-w_{-}$ is the hysteresis width. In our
case the dynamical exponent is $a=0$ near the transition (i.e.
there is a constant gap) and $a=1$ away from it, leading to
the values $\eta=1,1/2$ as reported. We note that this
argument is similar to the one used to explain the Kibble-Zurich mechanism \cite{Keesling2019}, with the difference that there one usually imagines
scanning the parameter up to the critical point, rather than passing
through it. It also agrees with contemporary works on hysteresis in
bi-stable systems in the low and high scanning regimes \cite{Chen2026}.

Notably, the numerical results shown in Fig. 4 of the main text indicate
a scaling exponent $\eta\approx0.58$ instead of the expected $\eta=1/2$
power law. To explain this discrepancy, we remember a fundamental assumption
taken while estimating the hysteresis width, namely that the $\langle\hat{J}_{z}\rangle$
dynamics are characterized by a single decay rate $\lambda\left(w\right)$,
as given in Eq. (\ref{eq: relaxation}). Looking back at the full Lindbladian spectrum in Fig. \ref{fig: Lindbaldian spectrum}, we see that
the real situation is more complicated; There are multiple eigenvalues
characterizing the dynamics, and there is no clear separation
between the first dissipative mode (equal to $\lambda\left(w\right)$)
and the higher modes. This means that for rates $r\gg\gamma$ there
is no well defined single rate at which the system decays, and so our calculation
of the adiabatic boundary (as depicted in Fig. 3 of the paper) is "perturbed" by the influence of higher modes. This explains the slightly deviated exponent 0.58.

\end{document}